\documentclass[11pt,a4paper]{article} 
\usepackage{jheppub,hyperref,mdwlist}
\usepackage{jheppub,hyperref,mdwlist}
\usepackage{amsmath}
\usepackage{graphicx}
\usepackage{subfigure}
\usepackage{amssymb}
\usepackage{xcolor}
\usepackage{multirow}
\usepackage{cancel}
\usepackage{color}
\usepackage{mathtools}
\usepackage{listings}
\usepackage{xcolor}
\usepackage{float}
\usepackage{slashed}
\usepackage{bm}
\usepackage{amsmath}
\usepackage{wrapfig}

\newcommand{\be}{\begin{equation}}
\newcommand{\ee}{\end{equation}}
\newcommand{\beq}{\begin{equation}}
\newcommand{\eeq}{\end{equation}}
\newcommand{\bea}{\begin{eqnarray}}
\newcommand{\eea}{\end{eqnarray}}
\definecolor{airforceblue}{rgb}{0.36, 0.54, 0.66}
\definecolor{steelblue}{rgb}{0.27, 0.51, 0.71}
\definecolor{amber}{rgb}{1.0, 0.49, 0.0}

\title{\boldmath A two-component dark matter model with $Z_2 \times Z_4$ symmetry}

\author{XinXin Qi,}
\author{Hao Sun}
\affiliation{Institute of Theoretical Physics, School of Physics, Dalian University of Technology, No.2 Linggong Road, Dalian, Liaoning, 116024, P.R.China }
\emailAdd{qxx@dlut.edu.cn}
\emailAdd{haosun@dlut.edu.cn}

\abstract{
We consider a two-component dark matter model with $Z_2 \times Z_4$ symmetry, where a singlet scalar $S$ and a Majorana fermion $\chi$ are introduced as dark matter candidates. We also introduce another singlet scalar $S_0$ with a non-zero vacuum expectation value to the SM so that the fermion dark matter can obtain mass after spontaneous symmetry breaking. We have a new Higgs boson in the model and in the case of the decoupling limit, the fermion dark matter production is only determined by $S$ and the new Higgs boson. The mass hierarchy of these new particles can make a difference in the reaction rate of dark matter annihilation processes, contributing to different viable parameter spaces for different mass orderings. We randomly scanned the parameter space with six various cases under relic density constraint and found that when  $\chi$ is the lightest among the dark sector, $\chi$ production is generated via the so-called forbidden channels. Moreover, we consider the combined limits arising from Higgs invisible decay, dark matter relic density and direct detection constraints. Within the chosen parameter space, direct detection results put the most stringent constraint, and we have a more flexible value for the scalar dark matter mass when the mass of $\chi$ is not smaller than the new Higgs boson mass.
}

\begin{document}
\maketitle
\flushbottom

\section{Introduction}
Astronomical experiments indicate that more than $80\%$ of the matter in our Universe is composed of dark matter (DM)\cite{Kolb:1990vq}, however, the microscopic origin of DM is still mysterious and remains one of the most important questions in physics.  One of the well-known scenarios for DM is weakly interacting massive particle (WIMP)\cite{Lee:1977ua,Gondolo:1990dk,Jungman:1995df}, where DM mass is assumed to be at  GeV to TeV scale. However, according to the DM direct detection experiments such as PandaX \cite{PandaX-4T:2021bab} and LZ \cite{LZ:2024zvo}, there is no evidence for  DM  at present, and WIMP models are facing serious challenges for the null results. Generally speaking,  the  WIMP models often demand large annihilation interaction to obtain the observed DM relic density but direct detection experiments constrain the couplings to be small. One possible solution to alleviate such tension is multi-component DM models \cite{Boehm:2003ha,Barger:2008jx,Zurek:2008qg,Profumo:2009tb,Liu:2011aa,Qi:2024zkr,Pandey:2017quk,Bhattacharya:2016ysw,Bhattacharya:2017fid,Bhattacharya:2022qck,Sakharov:1994pr,Khlopov:2021xnw}, which include more than one dark matter candidates, and the quantity to be compared against the direct detection limits provided by the experimental collaborations is the product of the fraction of dark matter times the respective scattering cross section instead. 

Multi-component dark matter models have been discussed for a long time,  such as $Z_5$ two-component scalar DM model \cite{Belanger:2020hyh}, $Z_7$ three-component scalar DM model \cite{Belanger:2022esk} and so on.  Among these models, dark matter particles are stabilized by additional discrete symmetry, where the visible sector and dark sector will carry different charges, and particles in the dark sector can contribute to new processes such as co-annihilation \cite{Baker:2015qna}, semi-annihilation\cite{Belanger:2014bga}, co-scattering \cite{Alguero:2022inz} as well as other conversion processes between dark matter. On the other hand, since one has two or more two types of DM particles in the model, which constitutes the observed DM relic density totally,   each component can be generated a via different production mechanism.  

In this work, we consider a two-component dark matter model under $Z_2 \times Z_4$ symmetry, where DM relic density is obtained via the Freeze-out mechanism \cite{Chiu:1966kg}. Concretely speaking,  we introduce a singlet scalar $S$ and a Majorana fermion $\chi$  as dark matter candidates to the SM.  Research about both singlet scalar and fermion as dark matter candidates can be found in \cite{Yaguna:2021rds,Yaguna:2023kyu,Bhattacharya:2018cgx}, and in this work, the bare mass term of $\chi$ is forbidden due to the  $Z_2 \times Z_4$ symmetry, 
another singlet scalar $S_0$  with non-zero vacuum expectation value is therefore introduced so that $\chi$ can obtain mass after spontaneously symmetry breaking. Moreover,  the mass hierarchy of these new particles can make a difference in the reaction rate of dark matter annihilation processes, which will contribute to different viable parameter spaces for different mass ordering, and we have six different cases for the possible mass hierarchy.
Particularly,  in the case of the decoupling limit, $\chi$ production is completely determined by new Higgs as well as $S$ and independent of SM particles. Similar cases in a two-component  
dark matter model where $\mathrm{DM_1}$ is equilibrium with the SM bath and $\mathrm{DM_2}$ is little connection with the SM particles can be found in the so-called ``pseudo-FIMP'' (pFIMP) models \cite{Bhattacharya:2022dco,Lahiri:2024rxc}, and in this paper we focus on the case of WIMPs instead.  On the other hand,  when $\chi$ is lightest among the dark sector, $\chi$ production is generated with the so-called ``Forbidden channels", which are kinetically forbidden at zero temperature. Discussion about ``Forbidden dark matter " can be found in \cite{Li:2023ewv,Konar:2021oye,Griest:1990kh,Yang:2022zlh,DAgnolo:2020mpt,Abe:2024mwa,Duan:2024urq}, 
and in this work, we analyze the viable parameter space of the model including the ``Forbidden dark matter" case from the point of theoretical constraint, Higgs invisible decay, relic density and direct detection constraints.

The paper is arranged as follows, in section.~\ref{sec:2}, we  give the two-component  dark matter model with $Z_2\times Z_4$ symmetry. In section.~\ref{sec:3}, we briefly discuss the 
theoretical constraint on the model. In section.~\ref{sec:4}, we discuss the dark matter phenomenology of the model including Higgs invisible decay, dark matter relic density as well as direct detection, and finally we summarize in the last section of the paper.
\section{Model description}\label{sec:2}
In this part, we consider a two-component dark matter model with $Z_2 \times Z_4$ symmetry by introducing two singlet scalars $S$ and $S_0$ as well one Majorana fermion $\chi$ to the SM, where $S$ and $\chi$ are dark matter candidates and $S_0$ owns non-zero vacuum expectation value  (vev) $v_0$, and the charges the particles in the model carrying are listed as follows:
\begin{table}[htbp]
\center
 \begin{tabular}{|l|r|}
 \hline
 Particle  & $Z_2 \times Z_4$ \\
 \hline
 $\mathrm{SM}$    & (1,1)\\
 \hline
 $S$     & (-1,1)\\
 \hline
 $S_0$ & (1,-1)\\
 \hline
 $\chi$ & (1,i)\\
 \hline
  \end{tabular}
  \caption{ The charges of the particles  under $Z_2\times Z_4$ symmetry.}
  \label{table1}
\end{table}\\
The new additional Lagrangian is therefore given as follows:
\begin{align}
 \mathcal{L}_{new}  &\supset \frac{1}{2}M_1^2 S^2 + \frac{1}{4}\lambda_{s}S^4-\frac{1}{2} \mu_0^2S_0^2+\frac{1}{4}\lambda_{0} S_0^4 - \mu_H^2|H|^2 +\lambda_{H}|H|^4    
  +\lambda_{dh}S^2|H|^2 + \lambda_{ds}S^2S_0^2 \notag\\
 +& \lambda_{sh}S_0^2|H|^2 + y_{sf}S_0\chi^{T}\chi
 \end{align}
 where $H$ is the SM Higgs doublet. Under unitarity gauge, $H$ and $S_0$ can be expressed with:
    \begin{equation}
H=\left(\begin{array}{c} 0 \\ \frac{v+h}{\sqrt{2}}\end{array} \right) \, , \quad
S_0=s_0+ v_0\, ,\quad
\end{equation}
 where $v =246$ GeV corresponds to the electroweak symmetry breaking vev and $v_0$ is the vev of $S_0$. After spontaneous symmetry breaking (SSB), the masses of $S$ and $\chi$ can be given by:
 \begin{eqnarray}
 m_S^2= M_1^2 +2\lambda_{ds}v_0^2 +\lambda_{dh}v^2, ~~m_{\chi}=y_{sf}v_0,
 \end{eqnarray}
  where $m_S(m_{\chi})$ represents the mass of $S(\chi)$. On the other hand, we have the squared mass matrix of $s_0$ and $h$ with:
  \begin{eqnarray}
    \mathcal{M}= \left(
    \begin{array}{cc}
     2\lambda_{0}v_0^2 & \lambda_{sh}vv_0 \\
     \lambda_{sh} vv_0 & 2\lambda_{H}v^2 \\
    \end{array}
    \right).
  \end{eqnarray}
The physical masses of the two Higgs states $h_1, h_2$ are then 
\begin{align}
\label{Higgsmass}
	m^2_{1} &= \lambda_H v^2 + \lambda_{0} v_0^2 
	- \sqrt{(\lambda_H v^2 - \lambda_{0} v_0^2)^2 + (\lambda_{sh}vv_0)^2},\notag\\
  m^2_{2} &= \lambda_H v^2 + \lambda_{0} v_0^2 
	+ \sqrt{(\lambda_H v^2 - \lambda_{0} v_0^2)^2 + (\lambda_{sh}vv_0)^2}
\end{align} 
The mass eigenstate ($h_1,h_2)$ and the gauge eigenstate ($h, s_0$) can be related via
\begin{align}
\label{Higgs mixing}
	\begin{pmatrix}
		h_1 \\ h_2
	\end{pmatrix} = 
	\begin{pmatrix}
    	\cos\theta & -\sin\theta \\
		\sin\theta &  \cos\theta
	\end{pmatrix}
	\begin{pmatrix}
		h\\ s_0
	\end{pmatrix}.
\end{align} 
where
 \begin{eqnarray}
    \tan 2\theta= \frac{\lambda_{sh}vv_0}{\lambda_{0}v_0^2 - \lambda_{H}v^2}
\end{eqnarray}
Furthermore, we can assume $h_1$ is the observed SM Higgs and $h_2$ is the new Higgs in our model. One can choose the masses of the Higgs particles $m_{1}$ and $m_{2}$ as the inputs so that the couplings of $\lambda_H$, $\lambda_{0}$ and $\lambda_{sh}$ can be given by:
\begin{align}
\label{para_quartic}
	\lambda_H &= 
 		\frac{(m_{1}^2 +m_{2}^2) - 
    	\cos 2 \theta (m_{2}^2 - m_{1}^2)}{4 v^2}, \nonumber\\
	\lambda_{0} &= 
 		\frac{(m_{1}^2 +m_{2}^2) + 
    	\cos 2 \theta (m_{2}^2 - m_{1}^2)}{4 v_0^2} , \\
	\lambda_{sh} &= 
 		\frac{\sin 2 \theta (m_{2}^2 - m_{1}^2)}{2 v v_0}  \nonumber
\end{align}
According to the current results, the mixing angle of the SM Higgs with other scalars is limited stringently arising from  W boson mass correction \cite{Lopez-Val:2014jva} at NLO, the requirement of perturbativity and unitarity of the theory \cite{Robens:2021rkl} as well as the LHC and LEP direct search \cite{CMS:2015hra,Strassler:2006ri}. In this work, we consider the decoupling limit with $\sin\theta \to 0$ so that dark matter $\chi$ production is dominated by the new Higgs $h_2$ and the scalar dark matter $S$, where the relevant SM production is highly suppressed due to the tiny $\sin\theta$.

                                                                                                                                                                                                                                                                                                                                                                                                                                                                                                                                                                                                                                                                                                                                                                                                                                                                                                                                                                                                                                                                                                                                                                                                                                                                                                                                                                                                                                                                                                                                                                                                                                                                                                                                                                                                                                                                                                                                                                                                                                                                                                                                                                                                                                                                                                                                                                                                                                                                                                                                                                                                                                                                                                                     \section{Theoretical constraint}\label{sec:3}
                                                                                                                                                                                                                                                                                                                                                                                                                                                                                                                                                                                                                                                                                                                                                                                                                                                                                                                                                                                                                                                                                                                                                                                                                                                                                                                                                                                                                                                                                                                                                                                                                                                                                                                                                                                                                                                                                                                                                                                                                                                                                                                                                                                                                                                                                                                                                                                                                                                                                                                                                                                                                                                                                                                    In this section, we discuss the theoretical constraints on the model from the point of perturbativity, unitarity perturbativity and vacuum stability.
                                                                                                                                                                                                                                                                                                                                                                                                                                                                                                                                                                                                                                                                                                                                                                                                                                                                                                                                                                                                                                                                                                                                                                                                                                                                                                                                                                                                                                                                                                                                                                                                                                                                                                                                                                                                                                                                                                                                                                                                                                                                                                                                                                                                                                                                                                                                                                                                                                                                                                                                                                                                                                                                                                                  
\subsection{perturbativity}
To ensure the perturbative model, the contribution
from loop correction should be smaller than the tree level
values, which put stringent constraints on the parameters with:
\begin{eqnarray}
|2\lambda_{dh}|<4\pi,|2\lambda_{ds}|<4\pi,|y_{sf}|<\sqrt{4\pi}.
\end{eqnarray}
\subsection{unitarity perturbativiy}
The unitarity conditions come from the tree-level scalar-scalar scattering matrix which is
dominated by the quartic contact interaction. The s-wave scattering amplitudes should
lie under the perturbative unitarity limit, given the requirement the eigenvalues of the
S-matrix $\mathcal{M}$ must be less than the unitarity bound given by $|\mathrm{Re}\mathcal{M}| < \frac{1}{2}$.
\subsection{vacuum stability}
To obtain a stable vacuum, the quartic couplings in the scalar potential should be constrained, In our model, the scalar potential quartic terms can be given with a symmetric $3 \times 3$ matrix as follows:
\begin{eqnarray}\label{m1}
  \mathcal{S}=\left(
  \begin{array}{ccc}
    \lambda_{0}&\lambda_{sh}&\lambda_{ds}\\
     \lambda_{sh}&\lambda_{H}&\lambda_{dh}\\
     \lambda_{ds}&\lambda_{dh}&\frac{1}{4}\lambda_{s}\\
  \end{array}
      \right).
\end{eqnarray}
According to the copositive criterial, the vacuum stability demands the quartic couplings with:
\begin{align}
&\lambda_0,\lambda_H,\lambda_s \geqslant 0, \lambda_{sh}+\sqrt{\lambda_0\lambda_H} \geqslant 0, \lambda_{ds} +\frac{1}{2}\sqrt{\lambda_0\lambda_s}\geqslant 0, \lambda_{dh}+\frac{1}{2}\sqrt{\lambda_H\lambda_s}\geqslant 0,\notag\\
&\frac{1}{2}\sqrt{\lambda_s}\lambda_{sh}+\lambda_{ds}\sqrt{H}+\lambda_{dh}\sqrt{\lambda_0}+\sqrt{2(\lambda_{sh}+\sqrt{\lambda_0\lambda_H})(\lambda_{ds}+\frac{1}{2}\sqrt{\lambda_0\lambda_s})(\lambda_{dh}+\frac{1}{2}\sqrt{\lambda_H\lambda_s})} \notag\\
&+\frac{1}{2}\sqrt{\lambda_0\lambda_H\lambda_s}\geqslant 0.
\end{align}
\section{Dark matter phenomenology}\label{sec:4}
There are two dark matter candidates with $S$ and $\chi$ in the model, and in this part we discuss the dark matter phenomenology of the model.
\subsection{Higgs invisible decay}
In this work, we assume the decoupling limit so that the decay of SM Higgs into new Higgs is highly suppressed if the channel is kinetically allowed. On the other hand, when the scalar DM mass $m_S$ is smaller than $m_1/2$, the measured Higgs invisible decay at the LHC will impose stringent constraints on the decay width of $\Gamma_{h_1 \to SS}$. The expression of $\Gamma_{h_1 \to SS}$ is given by \cite{Han:2015hda}:
\begin{eqnarray}
\Gamma_{h_1 \to SS}=\frac{\lambda_{dh}^2v^2}{32\pi m_1}\sqrt{1-\frac{4m_S^2}{m_1^2}}
\end{eqnarray}
The current constraint according to the LHC result is \cite{CMS:2018yfx} with:
\begin{eqnarray}
\Gamma_{h_1 \to SS} \leqslant 0.16\Gamma_h,
\end{eqnarray}
where the SM Higgs decay with $\Gamma_h \approx 4.15$ MeV.
\subsection{Relic density}
 The current observed dark matter relic density given by the Planck collaboration is $\Omega_{DM}h^2 = 0.1198 \pm 0.0012$ \cite{Planck:2018vyg}, and we consider dark matter production in our model to be generated with the ``Freeze-out" mechanism. Both $S$ and $\chi$ will contribute to dark matter relic density and the Boltzmann equations for the abundance  of $S$
  and $\chi$ are given as follows:
  \begin{eqnarray*} \nonumber
\frac{dY_S}{dx} &=& \frac{1}{3H}\frac{ds}{dx}[- \langle \sigma v \rangle^{SS\to XX}(Y_S^2-{\bar{Y_S}}^2)
    -  \langle \sigma v \rangle^{ SS \to \chi \chi}(Y_S^2-Y_{\chi}^2 \frac{\bar{Y_S}^2}{\bar{Y_{\chi}}^2})\\ \notag
                        &-& \langle \sigma v \rangle^{SS\to h_{1,2}h_{1,2}}(Y_S^2-{\bar{Y_S}}^2)
		        -\theta(m_i-2m_S)\frac{\Gamma_{h_iS}}{s}(Y_{S}-\bar{Y_{S}})]
\end{eqnarray*}
\begin{eqnarray}
 \frac{dY_{\chi}}{dx}  &=&  \frac{1}{3H}\frac{ds}{dx} [- \langle \sigma v \rangle^{\chi\chi\to h_2h_2}(Y_{\chi}^2-{\bar{Y_{\chi}}}^2)  -  \langle \sigma v \rangle^{ \chi\chi \to S S}(Y_{\chi}^2-Y_S^2 \frac{\bar{Y_{\chi}}^2}{\bar{Y_S}^2})\notag\\
 &+&\theta(m_2-2m_{\chi})\frac{\Gamma_{h_2\chi}}{s}(Y_{\chi}-\bar{Y_{\chi}})] \ \ \ \         
\end{eqnarray}
where $i=1,2$, $x=m_S/T$ with $T$ being temperature, $\theta(x)$ is the Heaviside function, $s$ denotes the entropy density, $Y_S$ and $Y_{\chi}$ are abundance of $S$ and $\chi$ defined by $Y_S \equiv n_S/s$ and $Y_{\chi} \equiv n_{\chi}/s$,
where $n_S$ and $n_{\chi}$ are number density of $S$ and $\chi$. $\bar{Y_S}$ and $\bar{Y_{\chi}}$ are the abundance in thermal equilibrium, which are defined by:
\begin{eqnarray}\label{be}
\bar{Y_S}(x)=\frac{45}{4\pi^4}\frac{x^2}{g_{*S}}K_2(x),
\bar{Y_{\chi}}=\frac{45x^2m_{\chi}^2}{2\pi^4 g_{*S}m_S^2}K_2(\frac{m_{\chi}}{m_S}x).
\end{eqnarray}
where $K_2(x)$ is the modified Bessel function of the second kind and $g_{*S}$ is the number  effective degrees of freedom.
$H$ is the Hubble expansion rate of the Universe, $X$ denotes SM particles and $\langle \sigma v \rangle$ is the thermally averaged annihilation cross section. $\Gamma_{h_{1,2}S}$ and $\Gamma_{{h_2}\chi}$  represent the 
thermally averged decay rate of $h_{1,2} \to SS$ and $h_2 \to \chi\chi$, which are defined by \cite{Zhang:2024sox}:
\begin{eqnarray}
\Gamma_{h_1S}=\Gamma_{h_1 \to SS}\frac{K_1(m_1/T)}{K_2(m_1/T)},\Gamma_{h_2S}=\Gamma_{h_2 \to SS}\frac{K_1(m_2/T)}{K_2(m_2/T)},
\Gamma_{h_2\chi}=\Gamma_{h_2 \to \chi\chi}\frac{K_1(m_2/T)}{K_2(m_2/T)}.
\end{eqnarray}
with
\begin{eqnarray}
\Gamma_{h_2 \to SS}=\frac{\lambda_{ds}^2v^2}{32\pi m_1}\sqrt{1-\frac{4m_S^2}{m_2^2}},
\Gamma_{h_2 \to \chi\chi}=\frac{y_{sf}^2m_2}{4\pi}(1-\frac{4m_{\chi}^2}{m_2^2})^{3/2},
\end{eqnarray}
 where $K_1(x)$ is the modified Bessel function of the first kind. The second terms in each equation of Eq.~\ref{be} correspond to the conversion between dark matter particles, which can be influenced by the mass hierarchy between $m_S$ and $m_{\chi}$.

 To calculate the DM relic density numerically we use the micrOMGEAs 6.0 package \cite{Alguero:2023zol}. In addition, the model has been implemented through the FeynRules package \cite{Alloul:2013bka}. Note that the mass hierarchy of $m_S$,$m_{\chi}$ and $m_2$ will affect the efficiency of the processes related to dark matter production, which demands different viable parameter spaces under dark matter relic density constraint. 
   \begin{figure}[htbp]
\centering
 \subfigure[]{\includegraphics[height=4cm,width=4.9cm]{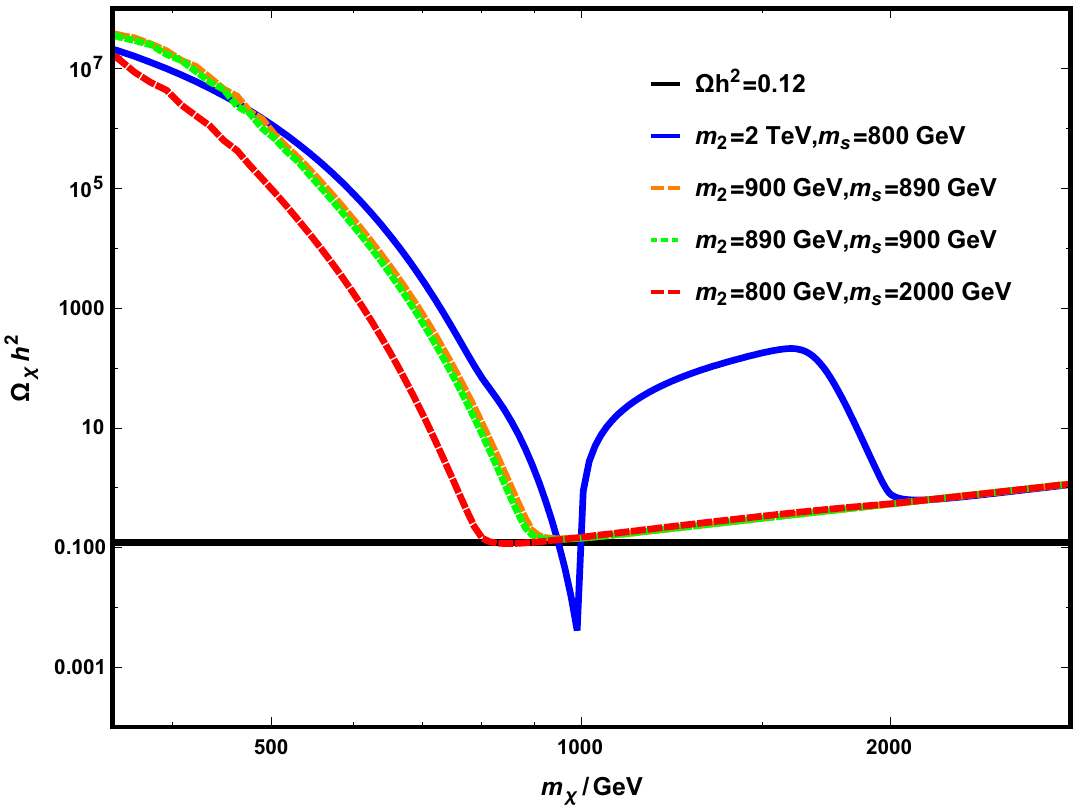}}
\subfigure[]{\includegraphics[height=4cm,width=4.9cm]{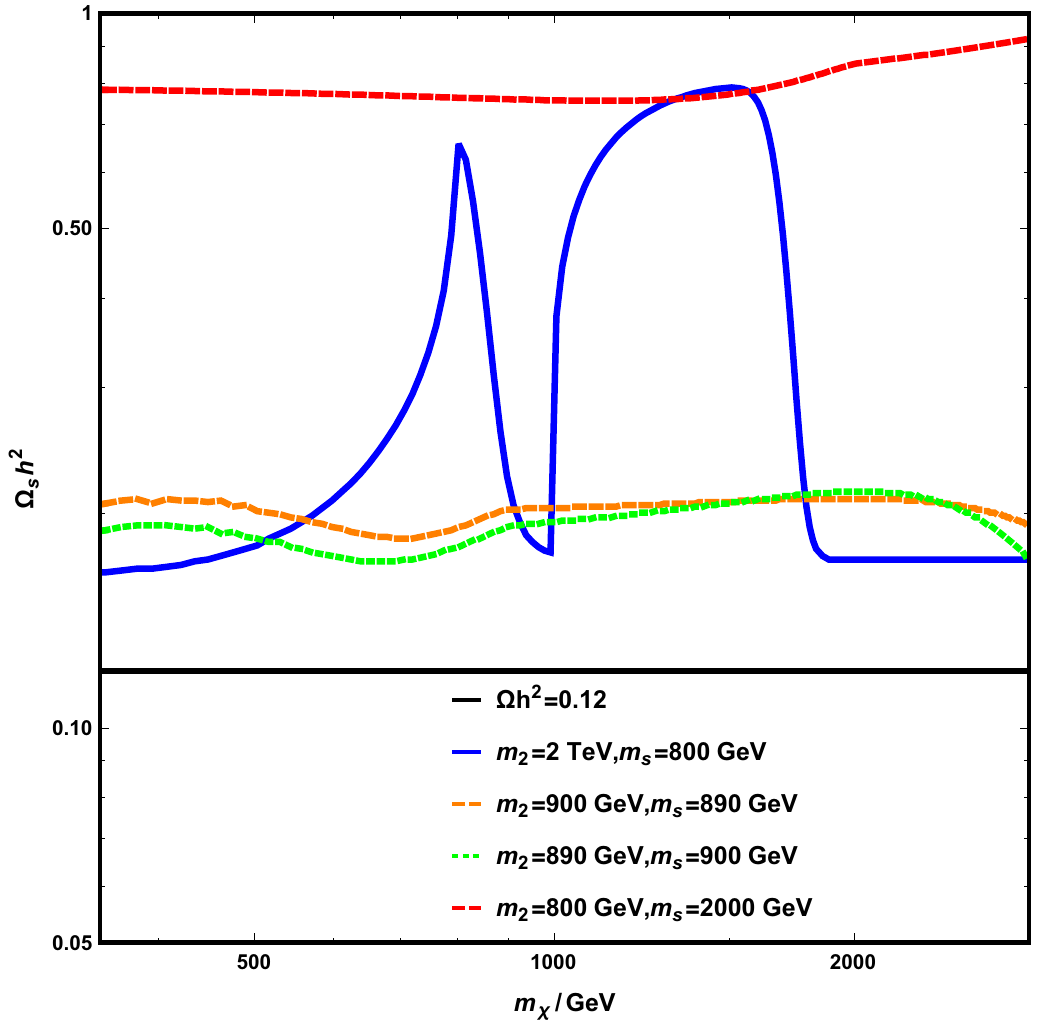}}
\caption{Evolution of $\Omega_{\chi}h^2$ (left) and $\Omega_{S}h^2$ (right) with $m_{\chi}$, where  we fixed $y_{sf}=0.1,\lambda_{ds}=0.1$ and $\lambda_{dh}=1$.
 The black lines correspond to the observed dark matter relic density value while other colored lines represent $(m_S,m_{\chi})$ taking different values. }
\label{fig1}
\end{figure}

 In Fig.~\ref{fig1}, we show the evolution of $\Omega_{\chi}h^2$ (left) and $\Omega_{S}h^2$ (right) with $m_{\chi}$, where  we fixed $y_{sf}=0.1,\lambda_{ds}=0.1$ and $\lambda_{dh}=1$.
 The black lines correspond to the observed dark matter relic density value while other colored lines represent $(m_S,m_{\chi})$ taking different values with $m_S > m_2$, $m_S < m_2$ and $m_S \sim m_2$.   For the four different cases, as we can see in Fig.~\ref{fig1}(a), $\Omega_{\chi}h^2$ decreases with the increase of $m_{\chi}$ when $m_{\chi}$ is small since the processes of $\chi\chi \to h_2h_2$ as well as  $\chi \chi \to SS$ are more efficient as $m_{\chi}$ becomes larger. The valleys in the four curves correspond to $m_{\chi} \approx m_2$, where the $\chi$-mediated t-channel processes are opened. Particularly, one can find a peak with $m_{\chi} \approx 1/2 m_2$ in the case of $m_2 =2$ TeV and $m_S=800$ GeV, where $\Omega_{\chi}h^2$ sharply decreases and interact with the black line (experiment result) arising from the Higgs-resonant effect. As $m_{\chi}$ becomes larger and eventually larger than $m_2$ and $m_S$, the four lines almost coincide with each other, which indicates the mass hierarchy between $m_2$ and $m_S$ makes little difference on $\Omega_{\chi}h^2$. According to Fig.~\ref{fig1}(b), although $m_{\chi}$ does not affect $\Omega_Sh^2$ directly, the mass hierarchy between $m_{\chi}$ and $m_S$ will influence the efficiency of the process of $\chi \chi \to SS$, which can play an important role in determining dark matter relic density as we can see the blue line in Fig.~\ref{fig1}(b) with $m_{\chi} \approx m_2/2$ (resonant-enhanced effect),$m_{\chi} \approx m_S$  and $m_{\chi} \approx m_2$ ($\chi$-mediated t-channel opened). 
 \subsection{Estimate on the parameters}
\begin{figure}[htbp]
\centering
 \subfigure[]{\includegraphics[height=4cm,width=4.9cm]{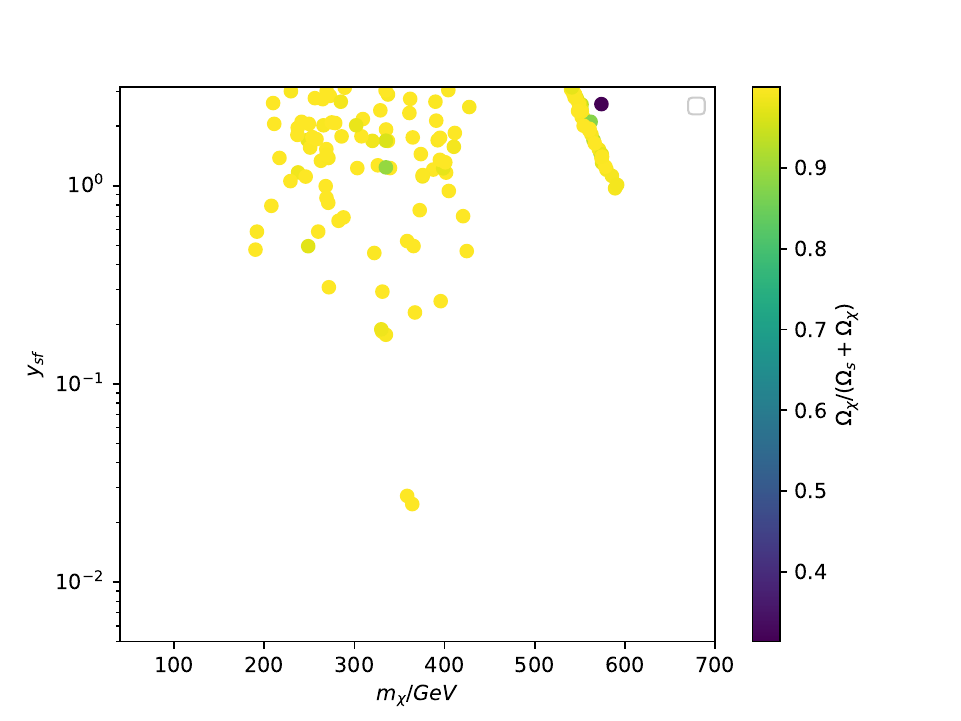}}
\subfigure[]{\includegraphics[height=4cm,width=4.9cm]{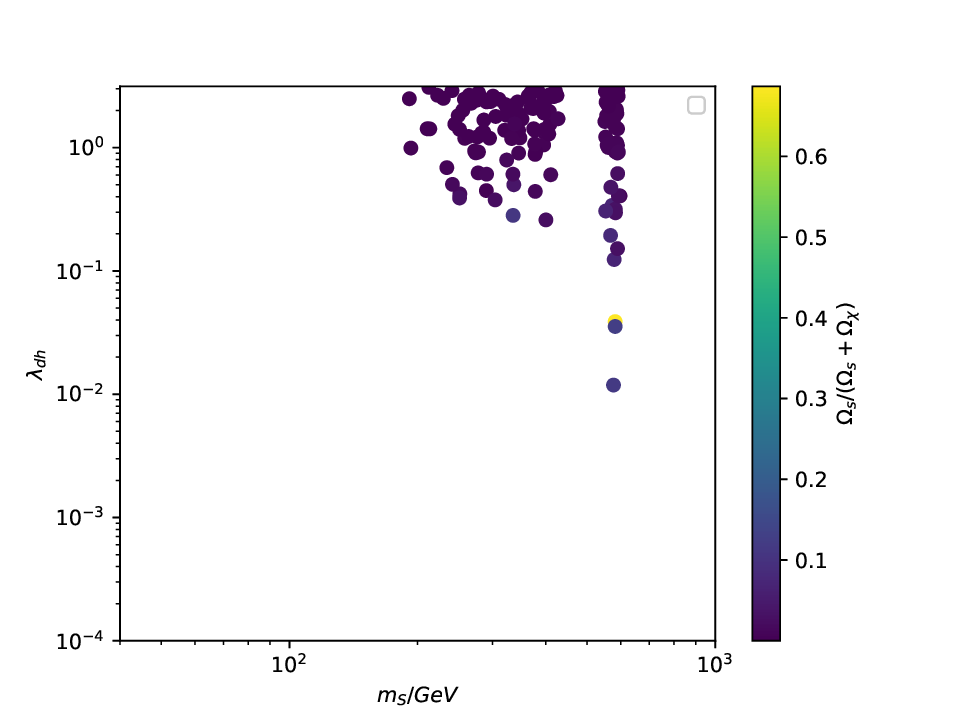}}
\subfigure[]{\includegraphics[height=4cm,width=4.9cm]{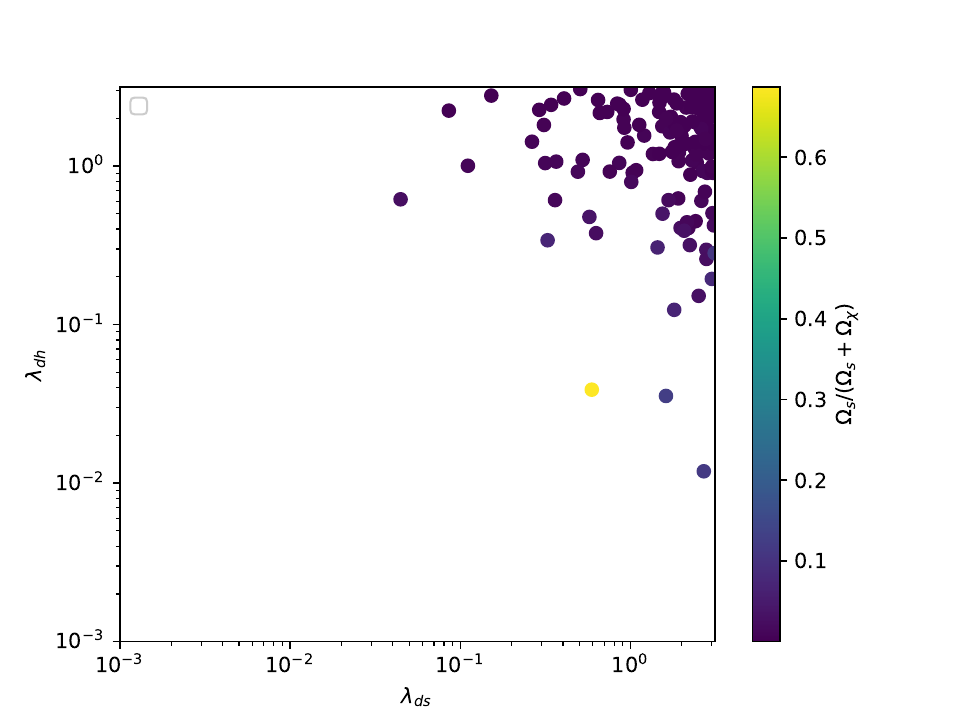}}
\caption{Viable parameter space of $m_{\chi}<m_S<m_2$, where points with different colors correspond to the fraction $\Omega_{\chi}/(\Omega_{\chi}+\Omega_S)$ in (a), $\Omega_{S}/(\Omega_{\chi}+\Omega_S)$ in (b) and (c) . }
\label{fig2}
\end{figure}

\begin{figure}[htbp]
\centering
 \subfigure[]{\includegraphics[height=4cm,width=4.9cm]{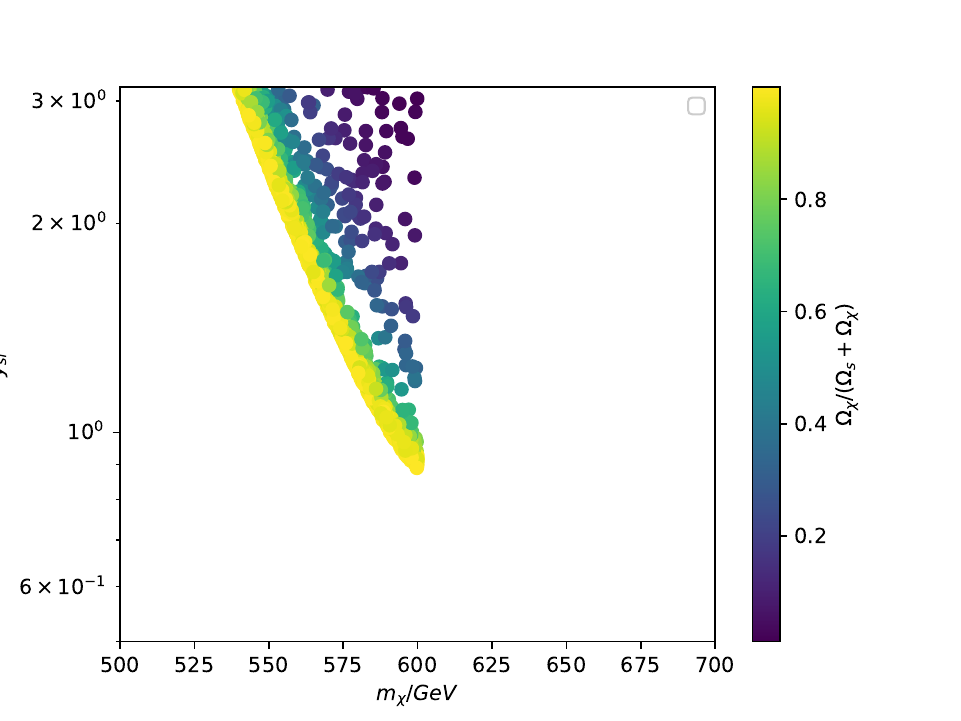}}
\subfigure[]{\includegraphics[height=4cm,width=4.9cm]{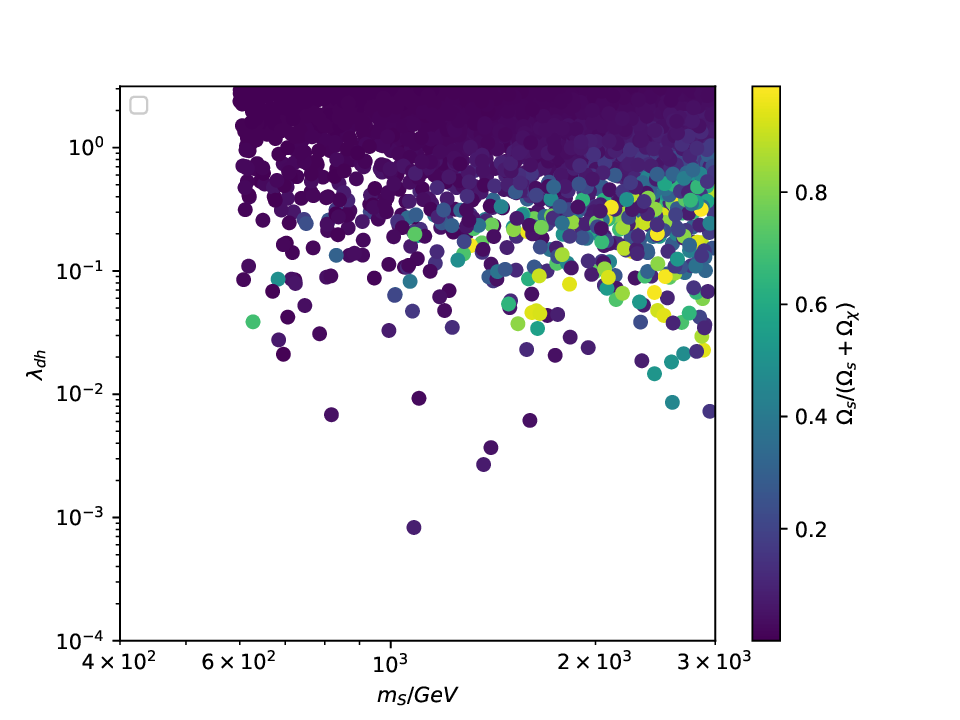}}
\subfigure[]{\includegraphics[height=4cm,width=4.9cm]{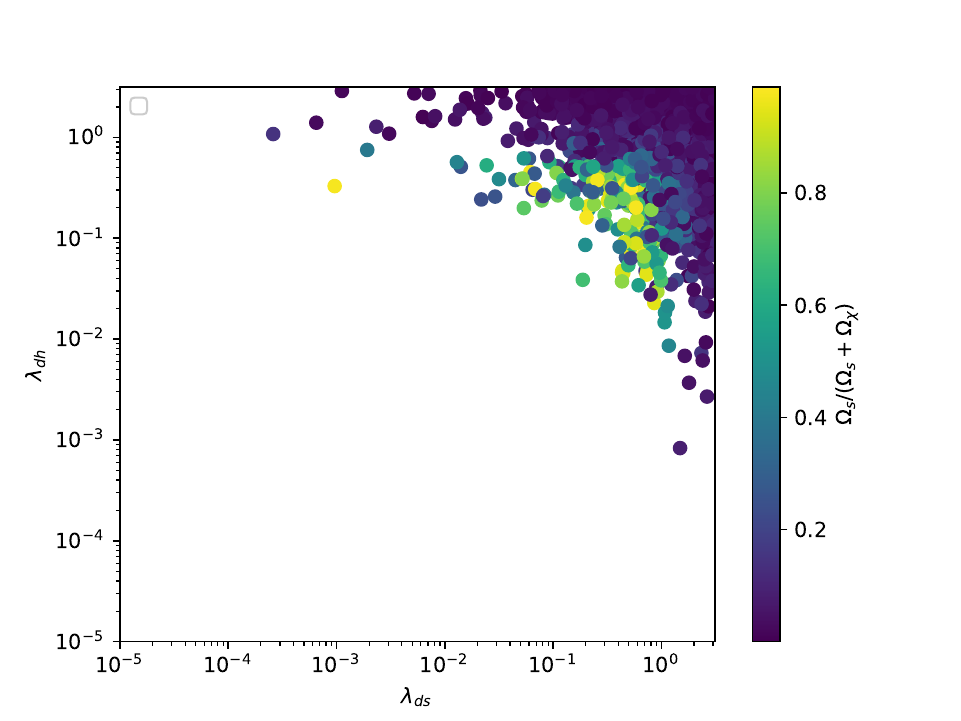}}
\caption{Viable parameter space of $m_{\chi}<m_2<m_S$, where points with different colors correspond to the fraction $\Omega_{\chi}/(\Omega_{\chi}+\Omega_S)$ in (a), $\Omega_{S}/(\Omega_{\chi}+\Omega_S)$ in (b) and (c) .}
\label{fig3}
\end{figure}

\begin{figure}[htbp]
\centering
 \subfigure[]{\includegraphics[height=4cm,width=4.9cm]{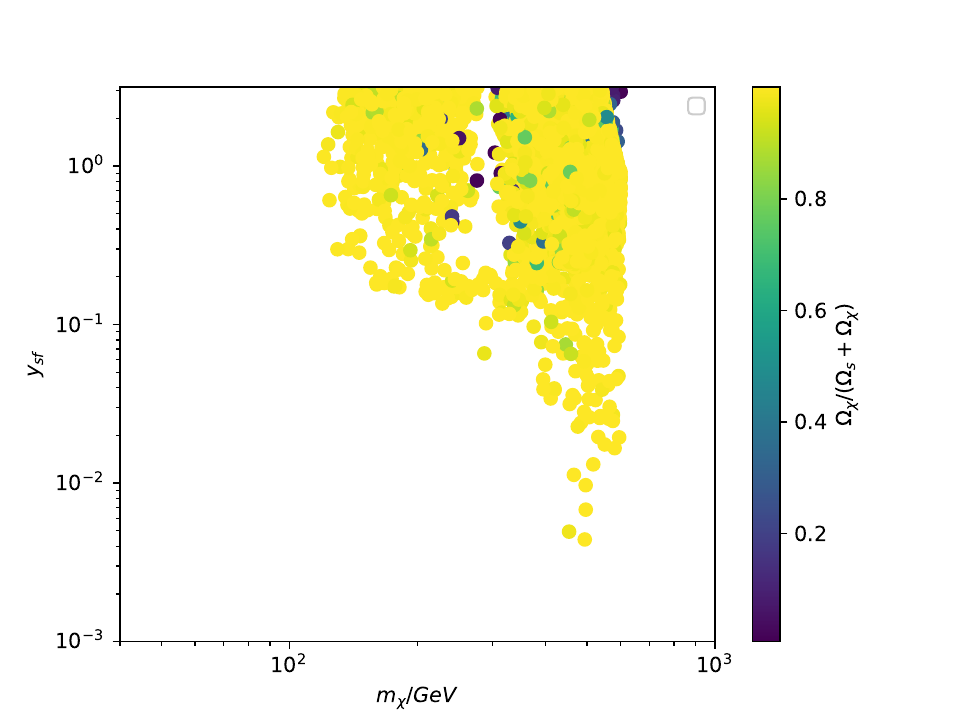}}
\subfigure[]{\includegraphics[height=4cm,width=4.9cm]{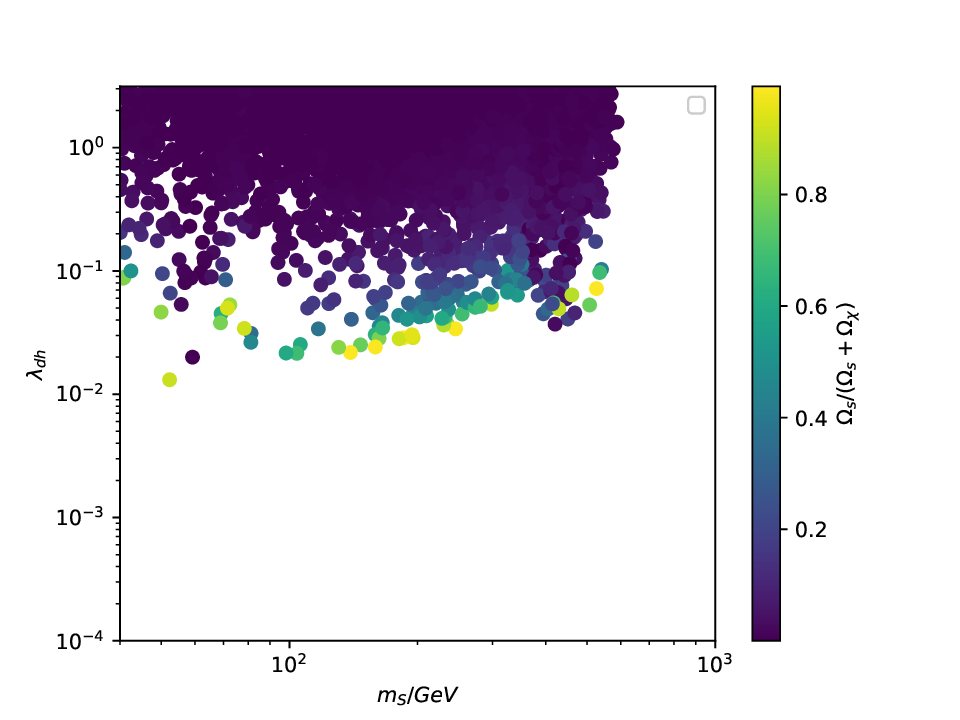}}
\subfigure[]{\includegraphics[height=4cm,width=4.9cm]{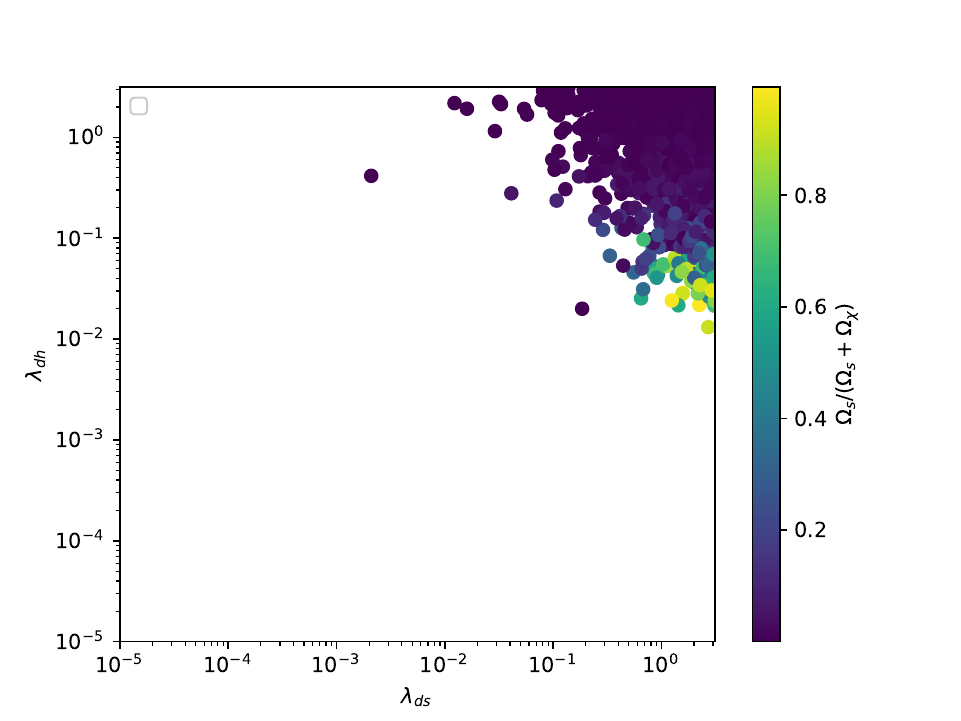}}
\caption{Viable parameter space of $m_S<m_{\chi}<m_2$, where points with different colors correspond to the fraction $\Omega_{\chi}/(\Omega_{\chi}+\Omega_S)$ in (a), $\Omega_{S}/(\Omega_{\chi}+\Omega_S)$ in (b) and (c) .}
\label{fig4}
\end{figure}

\begin{figure}[htbp]
\centering
 \subfigure[]{\includegraphics[height=4cm,width=4.9cm]{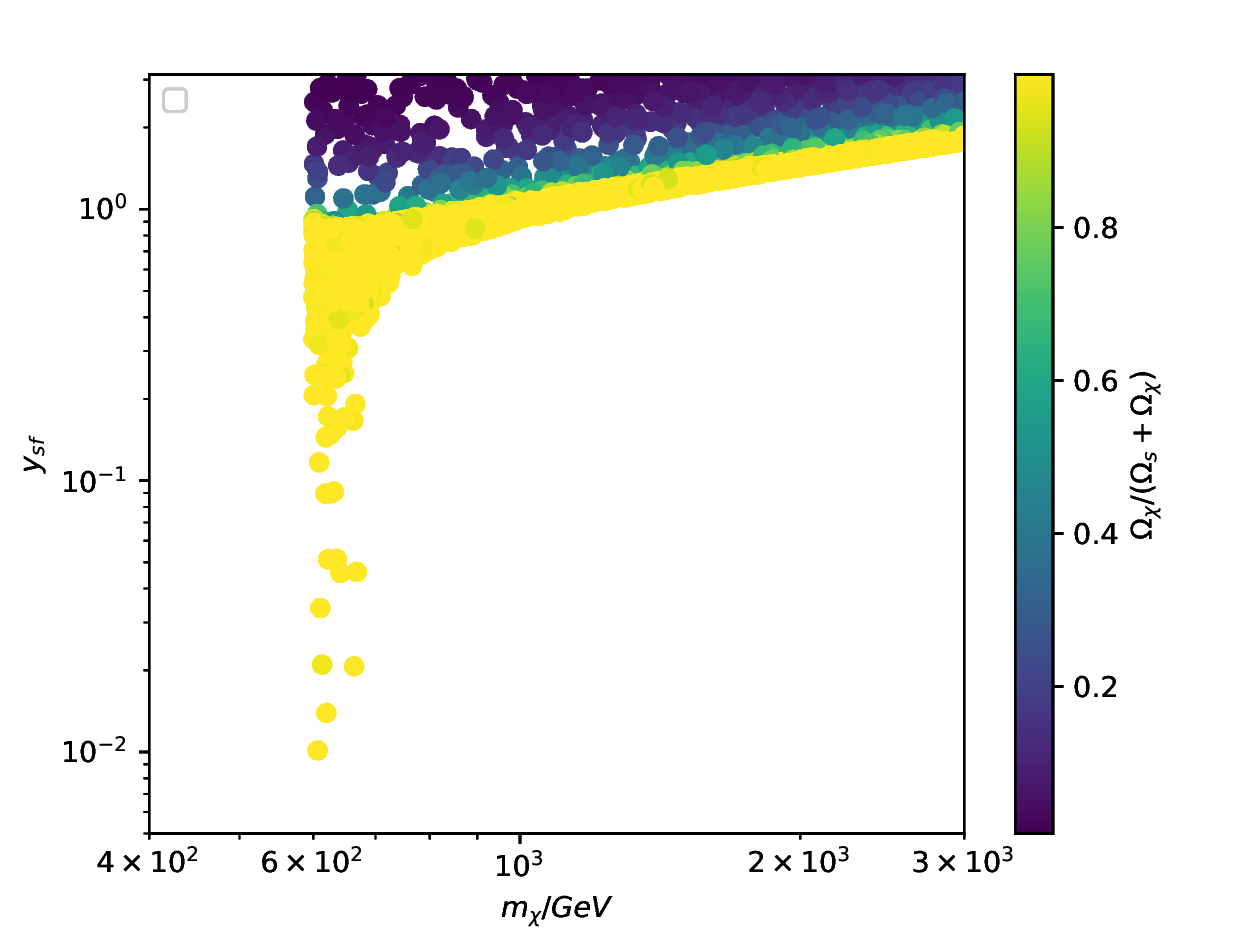}}
\subfigure[]{\includegraphics[height=4cm,width=4.9cm]{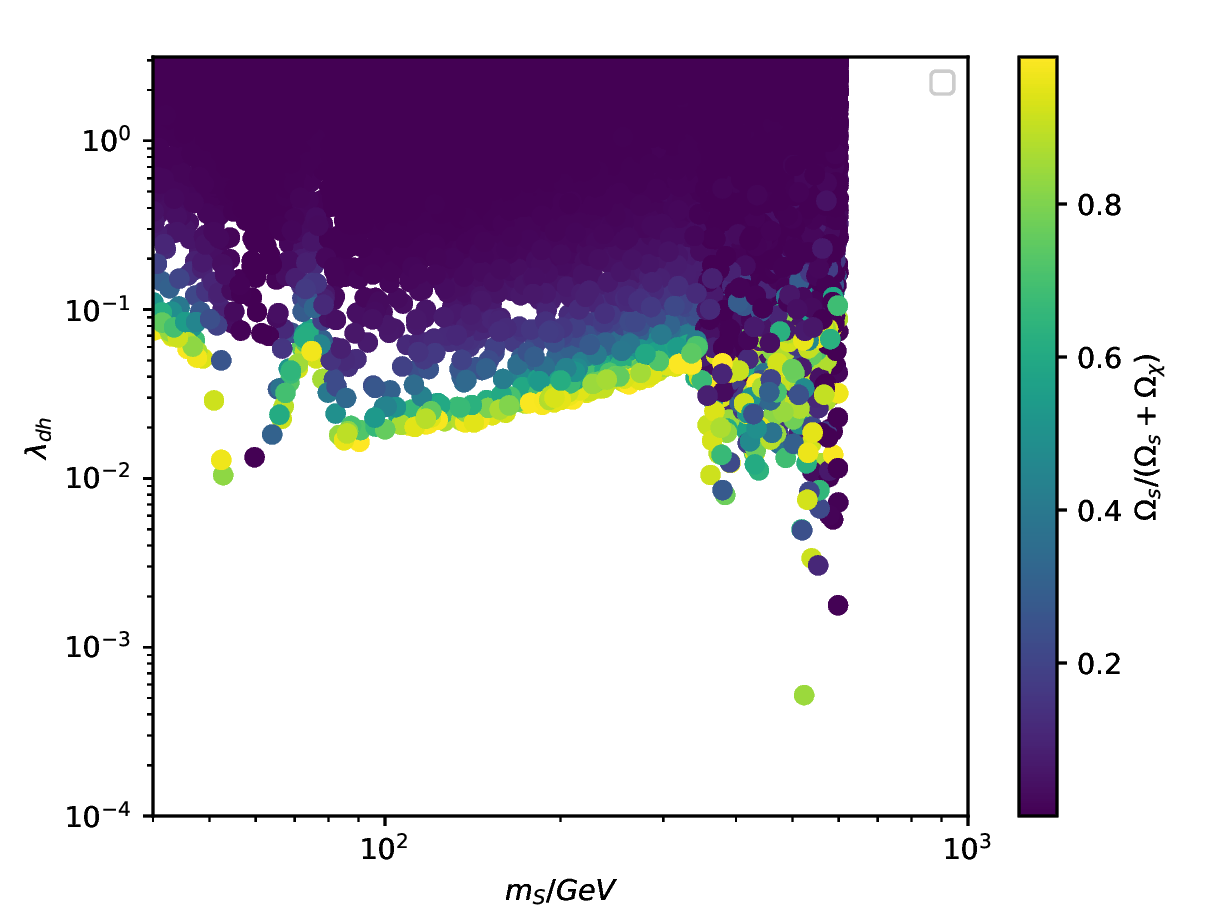}}
\subfigure[]{\includegraphics[height=4cm,width=4.9cm]{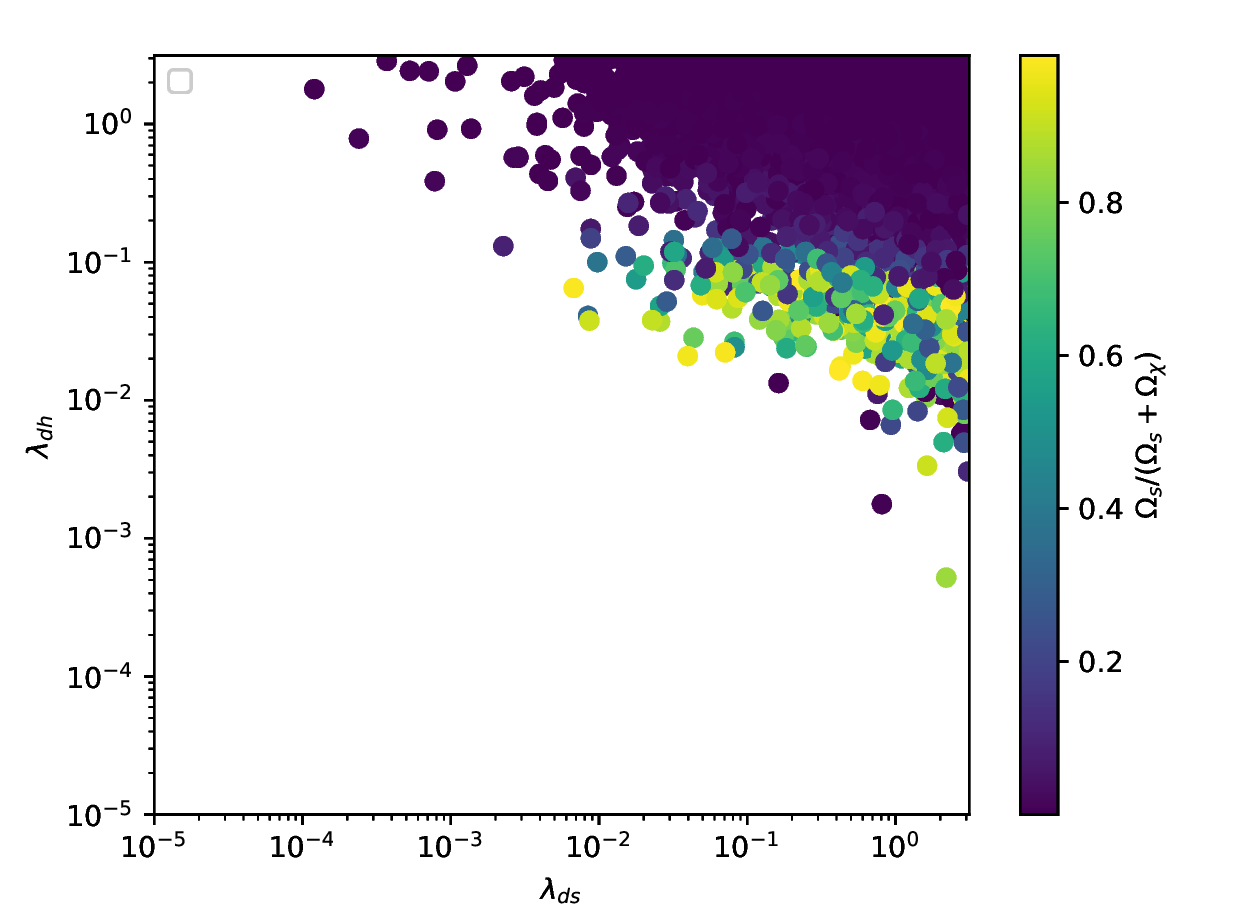}}
\caption{Viable parameter space of $m_S<m_2<m_{\chi}$,where points with different colors correspond to the fraction $\Omega_{\chi}/(\Omega_{\chi}+\Omega_S)$ in (a), $\Omega_{S}/(\Omega_{\chi}+\Omega_S)$ in (b) and (c) .}
\label{fig5}
\end{figure}

\begin{figure}[htbp]
\centering
 \subfigure[]{\includegraphics[height=4cm,width=4.9cm]{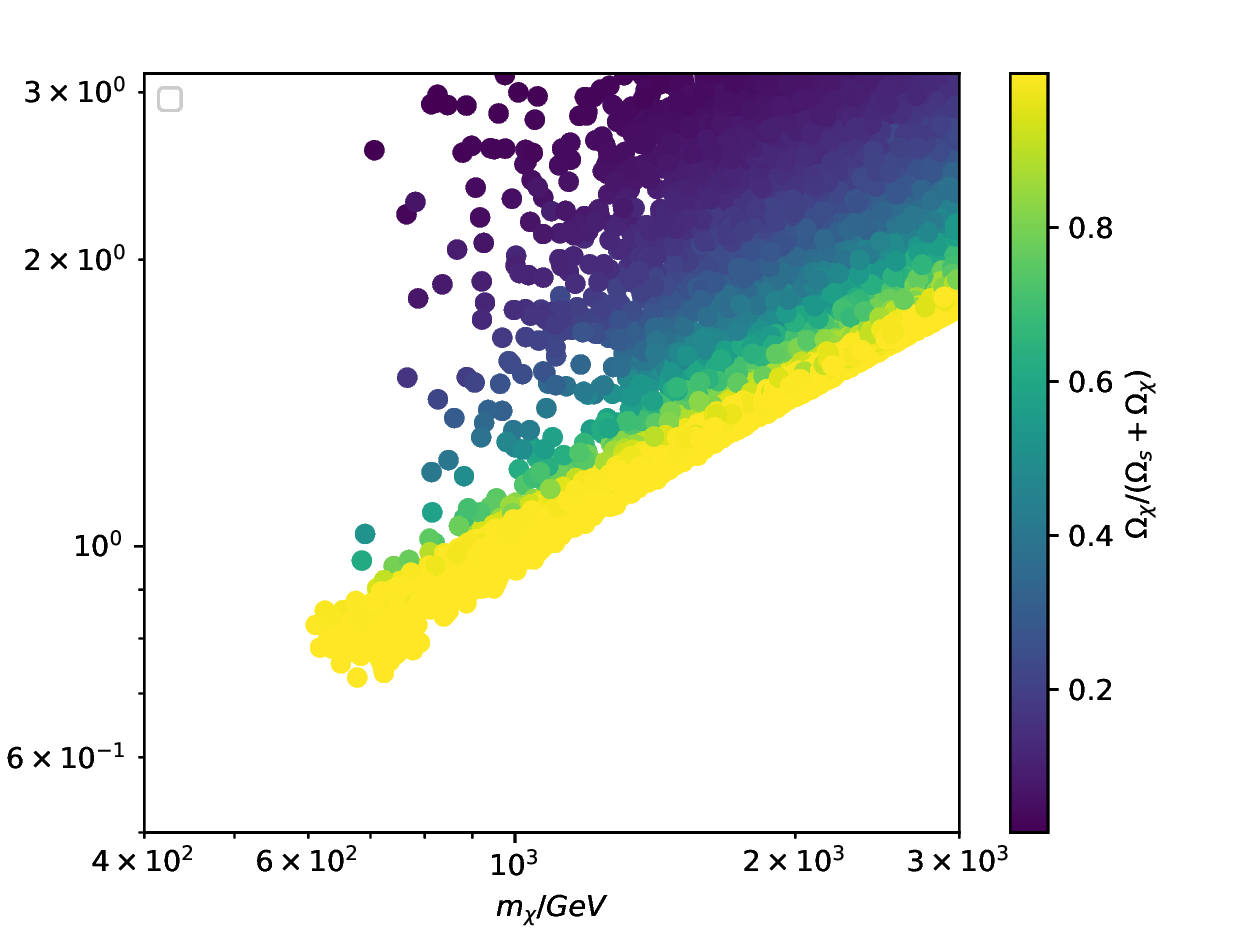}}
\subfigure[]{\includegraphics[height=4cm,width=4.9cm]{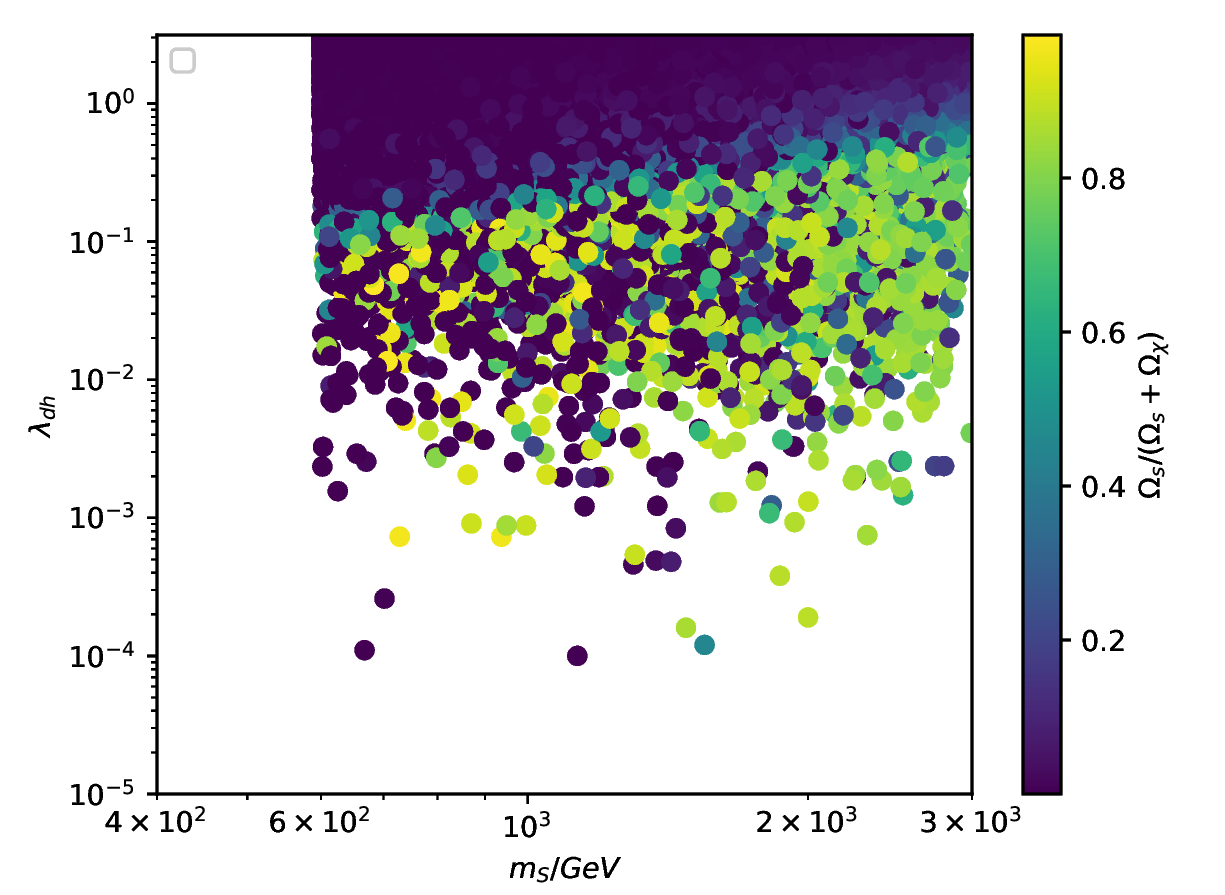}}
\subfigure[]{\includegraphics[height=4cm,width=4.9cm]{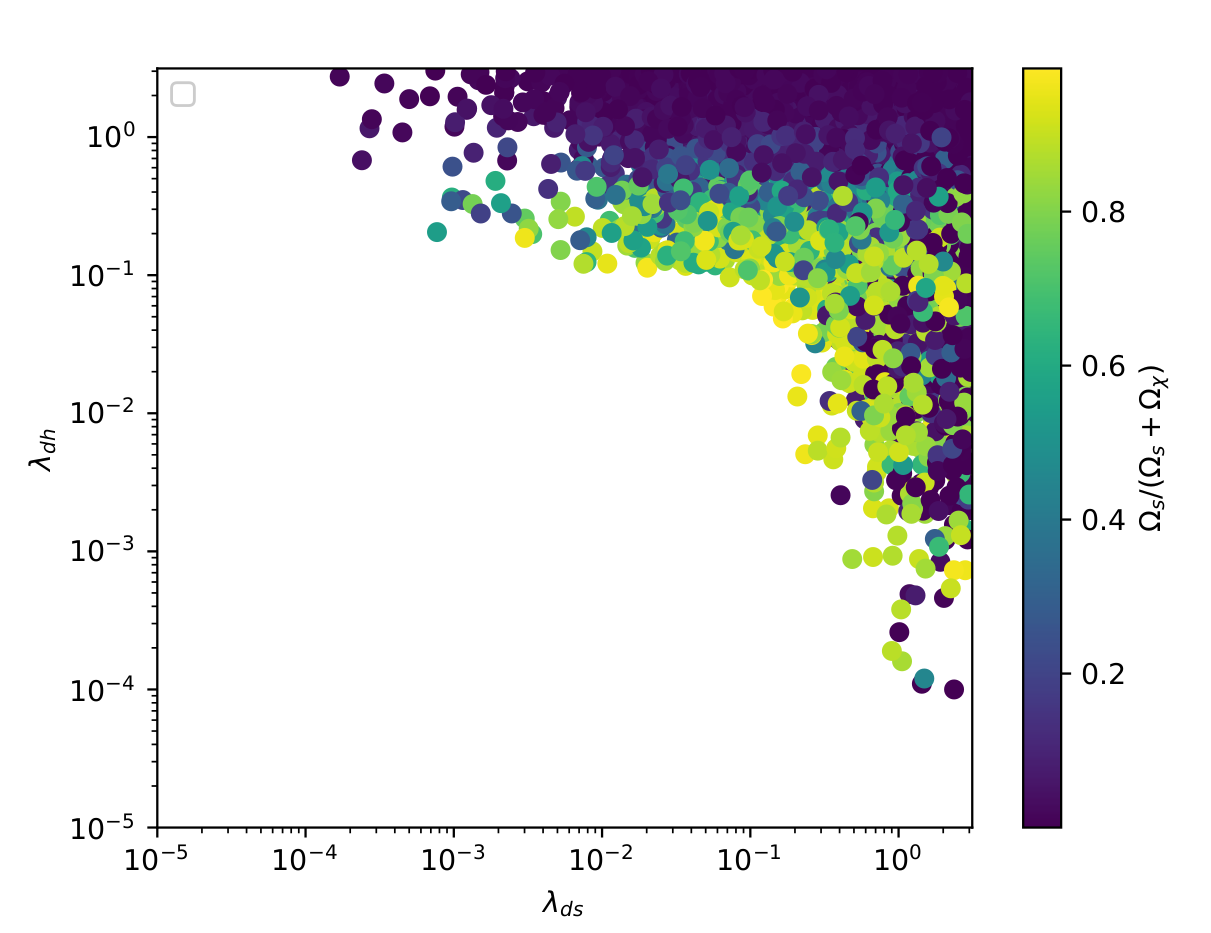}}
\caption{Viable parameter space of $m_2<m_S<m_{\chi}$, where points with different colors correspond to the fraction $\Omega_{\chi}/(\Omega_{\chi}+\Omega_S)$ in (a), $\Omega_{S}/(\Omega_{\chi}+\Omega_S)$ in (b) and (c) .}
\label{fig6}
\end{figure}

\begin{figure}[htbp]
\centering
 \subfigure[]{\includegraphics[height=4cm,width=4.9cm]{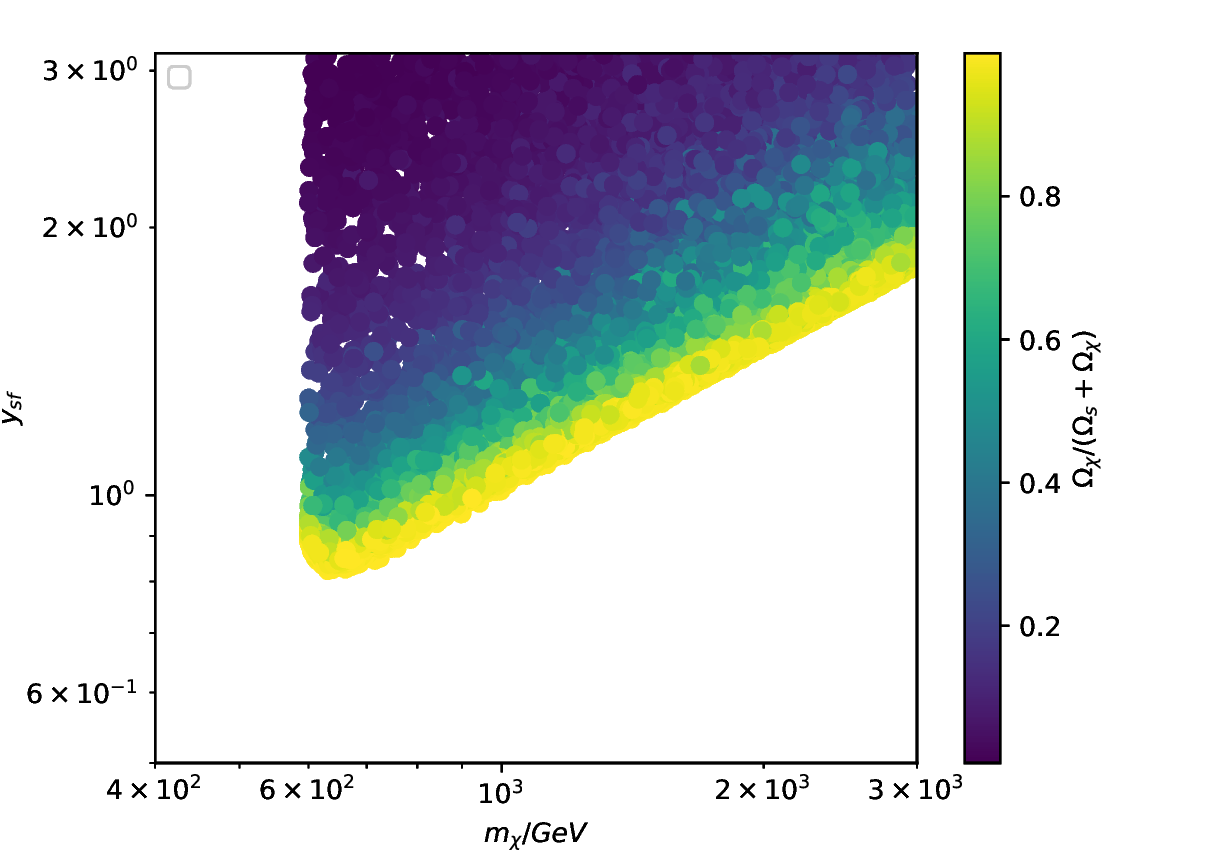}}
\subfigure[]{\includegraphics[height=4cm,width=4.9cm]{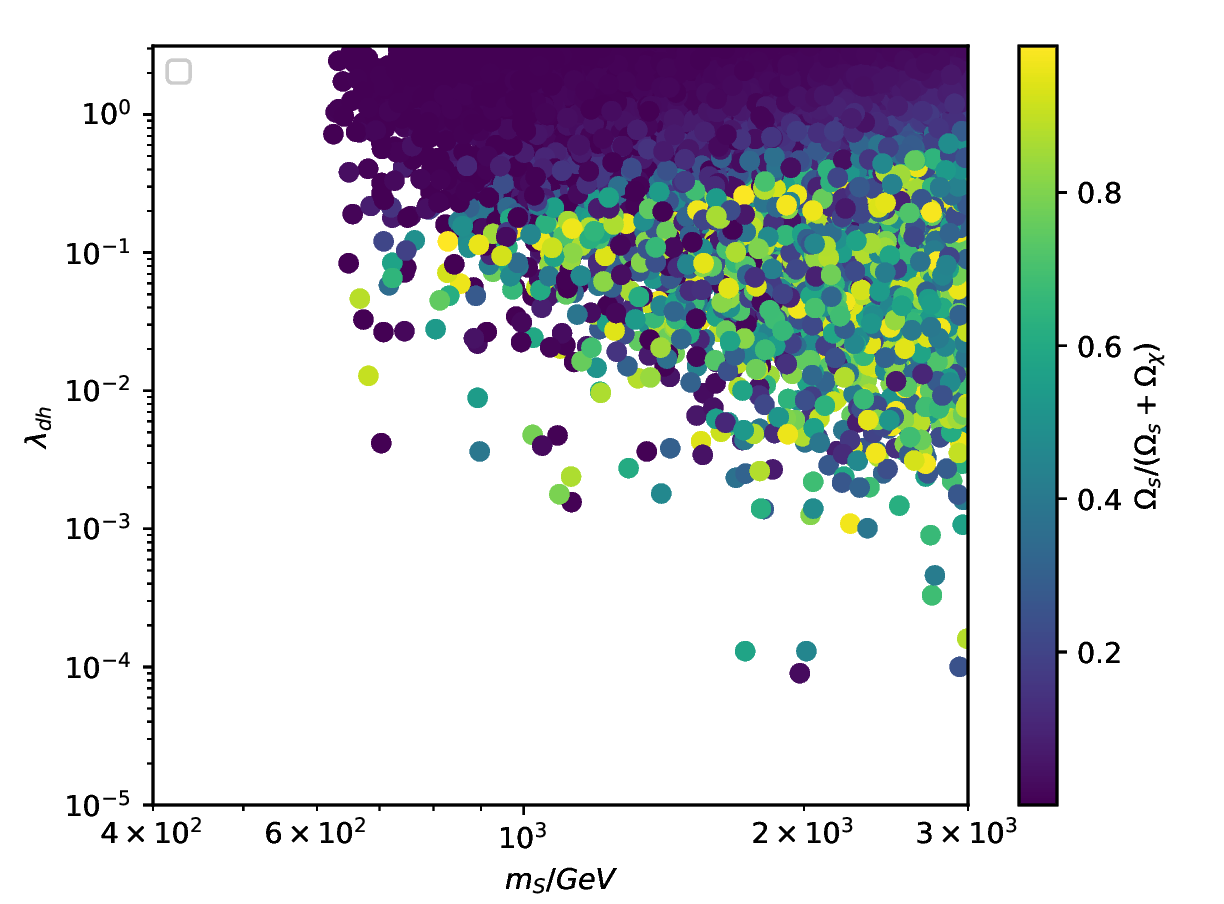}}
\subfigure[]{\includegraphics[height=4cm,width=4.9cm]{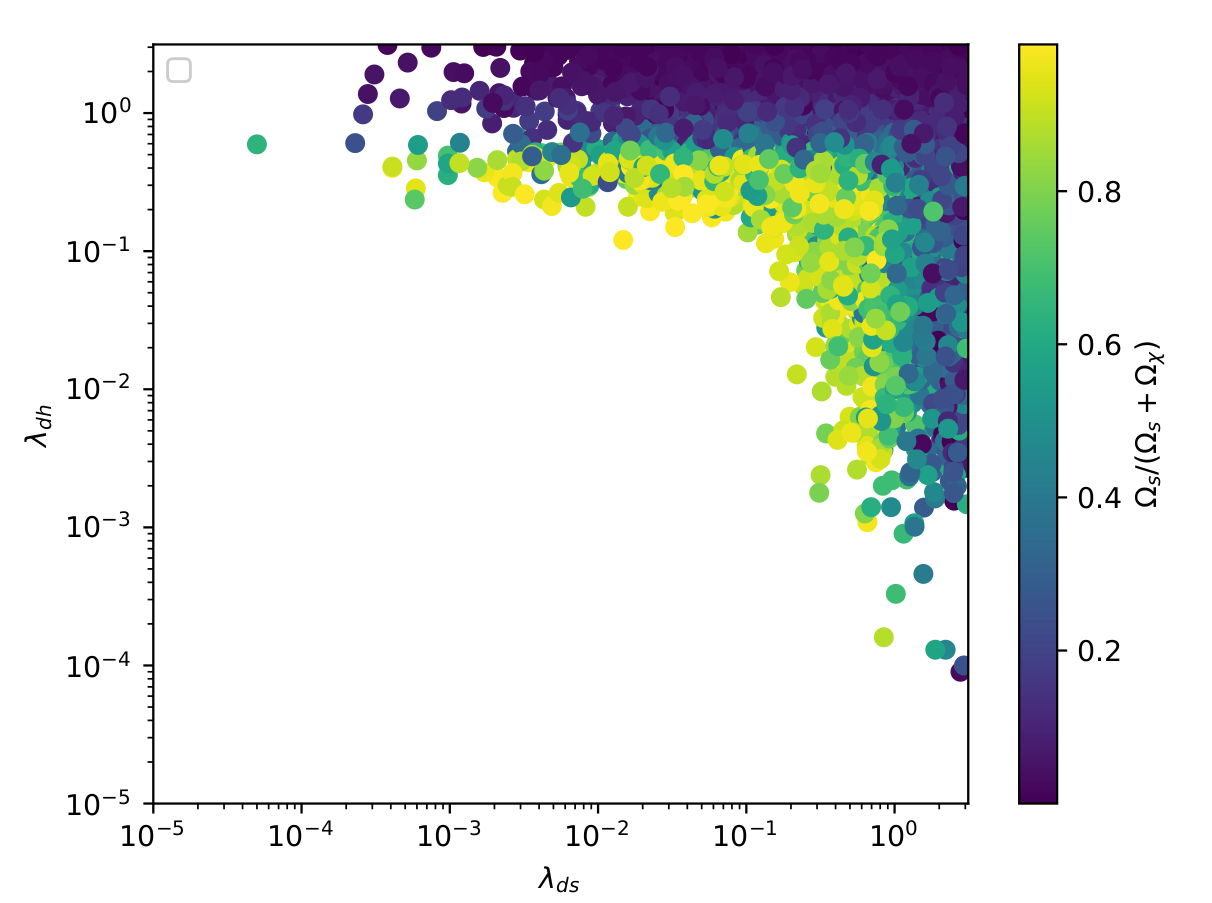}}
\caption{Viable parameter space of $m_2<m_{\chi}<m_S$, where points with different colors correspond to the fraction $\Omega_{\chi}/(\Omega_{\chi}+\Omega_S)$ in (a), $\Omega_{S}/(\Omega_{\chi}+\Omega_S)$ in (b) and (c) .}
\label{fig7}
\end{figure}

The mass hierarchy of $m_S,m_{\chi}$ and $m_2$ can make a difference in the evolution of dark matter as we show in Fig.~\ref{fig1}, which will contribute to different viable parameter spaces. Concretely speaking, we have six cases with $m_{\chi}<m_S<m_2$, $m_{\chi}<m_2<m_S$, $m_S<m_{\chi}<m_2$, $m_S<m_2<m_{\chi}$, $m_2<m_S<m_{\chi}$ and $m_2<m_{\chi}<m_S$. For simplicity, here we have omitted the cases of equal masses.
  To estimate the parameter space under the six cases, we make a random scan to consider the viable parameter space satisfying the dark matter relic density between 0.11 and 0.13, which amounts to about a $10\%$ uncertainty. The parameters are varied in the following ranges:
 \begin{eqnarray}
 m_{\chi},m_S \subseteq [40 ~\mathrm{GeV}, 3000 ~\mathrm{GeV}], \lambda_{ds},\lambda_{dh} \subseteq [10^{-5},3.14],
  y_{sf} \subseteq [0.001,3.14]
 \end{eqnarray}
 where we fixed $m_2= 600$ GeV, and we give the results of these six cases in Fig.~\ref{fig2} to Fig.~\ref{fig7}.
 
According to Fig.~\ref{fig2}, we set $m_{\chi}<m_S<m_2$ and points with different colors correspond to the fraction $\Omega_{\chi}/(\Omega_{\chi}+\Omega_S)$. Note that $m_{\chi}$ is the smallest among the dark sector, the annihilation channels of  $\chi\chi \to h_2h_2$ and $\chi\chi \to SS$ are kinetically forbidden at zero temperature but can proceed at finite temperature in the early universe, due to the thermal tail with high velocity $\chi$'s, where thermally averaged cross section for these channels are exponentially suppressed. It is easy for the density of $\chi$ to be over-abundant if these processes are not efficient enough, and the viable parameter space satisfying relic density constraint corresponds to the so-called ``forbidden dark matter" regime.
 In Fig.~\ref{fig2}(a), we show the viable parameter space of $m_{\chi}-y_{sf}$ satisfying relic density constraint. We have two separate regions for $m_{\chi}$ with $ 200~ \mathrm{GeV} \leqslant m_{\chi}\leqslant 400$ GeV and $ 500 ~\mathrm{GeV} \leqslant m_{\chi} \leqslant 600$ GeV. The former region corresponds to $m_{\chi} \approx m_S$ where the allowed value of $m_S$ is similar with $m_{\chi}$ according to Fig.~\ref{fig2}(b), and $\chi$ relic density is only determined by the forbidden channel $\chi \chi \to SS$, and $\chi$ is the dominant component among the dark matter production. Moreover, in the case of $m_{\chi} \approx m_2/2$, the allowed value of $y_{sf}$ can decrease to about 0.02 due to the Higgs-resonant effect. For $ 500 ~\mathrm{GeV} \leqslant m_{\chi} \leqslant 600$ GeV, the forbidden process of $\chi \chi \to h_2 h_2$ becomes efficient, which includes s-channel annihilation as well as $\chi$-mediated t-channel annihilation, and the density of $\chi$ can be much lower so that $S$ will be dominant in the relic density respectively depending on the interaction strength. According to Fig.~\ref{fig2}(b), the allowed value of $m_S$ is also divided into two regions $m_S \approx m_{\chi}$ and $m_S \approx m_2$.
 Although $S$ can still annihilate into SM particles regardless of the mass hierarchy, $\chi$ will be over-abundant if $S$ is much heavier than $\chi$ and $\chi\chi \to SS$ is not so efficient. For $m_S \approx m_{\chi}$, the allowed value for $\lambda_{dh}$ is limited to be larger than about 0.2 to guarantee a large annihilation cross-section under relic density constraint. 
 
 On the other hand, the allowed value for $\lambda_{dh}$ is more flexible in the case of $m_S \approx m_2$ since more channels are opened. For $\lambda_{ds}$, as we can see Fig.~\ref{fig2}(c), most of the points lie in the upper-right region and the small $\lambda_{ds}$ is excluded by the relic density constraint.
 
 In Fig.~\ref{fig3}, we show the viable parameter space of $m_{\chi}<m_2<m_S$. In this case, the channel $\chi \chi \to SS$ is highly suppressed for the large mass hierarchy between $m_S$ and $m_{\chi}$ when $m_{\chi}$ is small. Therefore, the viable parameter space of $m_{\chi}$ is limited within a narrow region with $ 525~ \mathrm{GeV} \leqslant m_{\chi} \leqslant 600$ GeV according to Fig.~\ref{fig3}(a), and the smaller $m_{\chi}$ is excluded for dark matter production being over-abundant. The allowed value for $y_{sf}$ is about $0.8\leqslant y_{sf} \leqslant 3.14$, and the lower bound of $y_{sf}$ decreases with the increase of $m_{\chi}$ contrary to the numerical relationship between $m_{\chi}$ and $y_{sf}$ instead, which indicates that relic density constraint puts a stringent limit on the parameter space. For a fixed $m_{\chi}$, a larger $y_{sf}$ will induce larger interaction strength so that the fraction $\Omega_{\chi}/(\Omega_{\chi}+\Omega_{S})$ will be smaller. For $m_S$, the allowed parameter space is much flexible with $600 \mathrm{GeV} \leqslant m_S \leqslant 3000$ GeV according to Fig.~\ref{fig3}(b), and most of points correspond to the small fraction of $\Omega_{S}/(\Omega_{\chi}+\Omega_{S})$. Particularly, when $\lambda_{dh}$ is larger than about 0.6, 
the annihilation of a pair of $S$ is so efficient and $S$ is always sub-dominant in dark matter relic density. In Fig.~\ref{fig3}(c), we show the viable parameter space of $(\lambda_{ds},\lambda_{dh})$, where all points lie in the upper-right region, and a small $\lambda_{ds}$ always demands a large $\lambda_{dh}$ under relic density constraint. Similarly, when $\lambda_{ds}$ is larger than about 1, we have a small fraction of $\Omega_{S}/(\Omega_{\chi}+\Omega_{S})$.

According to Fig.~\ref{fig4}, we give the results of  $m_S<m_{\chi}<m_2$. In this case, conversion of $\chi$ to S becomes more efficient via the annihilation process $\chi \chi \to SS$, the allowed value for $m_{\chi}$ is hence more flexible with about $100~ \mathrm{GeV} \leqslant m_{\chi} \leqslant 600$ GeV as we can see in  Fig.~\ref{fig4}(a).  We have  $0.004 \leqslant y_{sf} \leqslant 3.14$ under relic density constraint and when $m_{\chi} \approx m_2$, the value of $y_{sf}$ can decrease to 0.004 where the forbidden channel of $\chi \chi \to h_2 h_2$ is efficient. According to Fig.~\ref{fig4}(b), $m_S$ can take values ranging from [40 GeV, 600 GeV] and the viable parameter space of $\lambda_{dh}$ is about $0.01< \lambda_{dh} \leqslant 3.14$. With the increase of $\lambda_{dh}$, the fraction $\Omega_{S}/(\Omega_{\chi}+\Omega_{S})$ becomes smaller due to the larger annihilation cross-section, and for $\lambda_{dh} \geqslant 0.1$, $\chi$ is dominant in dark matter relic density. For $\lambda_{ds}$, we have similar conclusion with the case $m_{\chi}<m_2<m_S$ with $0.001 \leqslant\lambda_{ds} \leqslant 3.14$ according to Fig.~\ref{fig4}(c),and $S$ will be sub-dominant in dark matter production as long as $\lambda_{dh} \geqslant 0.1$  regardless of $\lambda_{ds}$ as we mentioned above.

 In Fig.~\ref{fig5}, we show the viable parameter space in the case of $m_S <m_2<m_{\chi}$, where the processes $\chi \chi \to SS$ and $\chi \chi \to h_2 h_2$ are kinetically allowed at zero temperature. The viable parameter space for $m_{\chi}$ is $ 600 ~\mathrm{GeV} \leqslant m_{\chi} \leqslant 3000$ GeV. With the increase of $m_{\chi}$, a larger $y_{sf}$  is demanded under relic density constraint, and for $y_{sf} \geqslant 1$, dark matter relic density is mainly determined by $S$. On the other hand, when $m_{\chi} \approx 600$ GeV,  we have a wider parameter space for $y_{sf}$ with $0.01 \leqslant y_{sf} \leqslant 3.14$ as we can see in Fig.~\ref{fig5}(a). The viable parameter space for $m_S$ is $40 ~\mathrm{GeV} \leqslant m_S \leqslant 600$ GeV with $0.0005 \leqslant \lambda_{dh} \leqslant 3,14$, and parameter space of $\lambda_{dh}$ is more flexible for the forbidden channels of $SS \to h_2h_2$ when $m_S \approx 600$ GeV according to Fig.~\ref{fig5}(b). Compared with the case of $m_S<m_{\chi}<m_2$, we have a wider parameter space for $(\lambda_{ds},\lambda_{dh})$ for the heavy $m_{\chi}$  as we can see in Fig.~\ref{fig5}(c).
 
 In Fig.~\ref{fig6} and Fig.~\ref{fig7}, we give viable parameter spaces of $m_2<m_S<m_{\chi}$ and $m_2<m_{\chi}<m_S$, where both $S$ and $\chi$ can annihilate into $h_2$. The value of $y_{sf}$ is limited within about [0.8,3.14] and the smaller $y_{sf}$ is excluded for dark matter being over-abundant. Compared with the former cases, there is a little difference between the results of $m_2<m_S<m_{\chi}$ and $m_2<m_{\chi}<m_S$  as $h_2$ is the lightest among the dark sector, and $\chi\chi \to h_2h_2$ as well as $SS \to h_2h_2$ are both kinetically allowed at zero temperature.
 
 As a summary, when $\chi $ is the lightest among the dark sectors, relic density of $\chi$ is determined by forbidden channels where $m_{\chi}$ is constrained within a narrow region. In the case of $S$ being the lightest,  $S$ can still annihilate into SM particles besides from the forbidden channels, and the process $\chi\chi \to SS$ is kinetically allowed at zero temperature. Therefore, we have a wider parameter space for $m_{\chi}$ and $m_S$. Furthermore, when $h_2$ is the lightest, the viable parameter spaces are more flexible as we mentioned above.
\subsection{Direct detection}
The Higgs portal interactions $\lambda_{dh}$  can contribute to the elastic scattering of the dark matter off nuclei in the model, which can put a stringent constraint on the parameter space. The expression of the spin-independent (SI) cross section can be given as follows\cite{Qi:2024uiz}:
  \begin{eqnarray}
  \sigma^{SI}= \frac{\lambda_{dh}^2}{4\pi}\frac{\mu_R^2m_p^2f_p^2}{m_H^4m_S^2}
    \label{ddeq}
  \end{eqnarray}
  where  $\mu_R$ is the reduced mass, $m_p$ is the proton mass, $m_H$ the SM Higgs mass and $f_p \approx 0.3$ is the quark content of the proton. Current experiments on the direct detection of dark matter can be found in \cite{PandaX-4T:2021bab,LZ:2024zvo}, and the LZ experiments \cite{LZ:2024zvo} put the most stringent constraint on the spin-independent dark matter. Since we have two dark matter particles but only $S$ can contribute to the elastic scatterings, the quantity to be compared
against the direct detection limits provided by the experimental collaborations is not the
cross-section itself but rather the product $\xi_S \sigma^{SI}$ with $\xi_S= \frac{\Omega_S}{\Omega_S+\Omega_{\chi}}$. Direct detection will also constrain the parameter space, and in the following discussion, the results are limited by both relic density constraint and direct direction constraint.
\subsection{Combined  results}
\begin{figure}[htbp]
\centering
 \subfigure[]{\includegraphics[height=5.5cm,width=6cm]{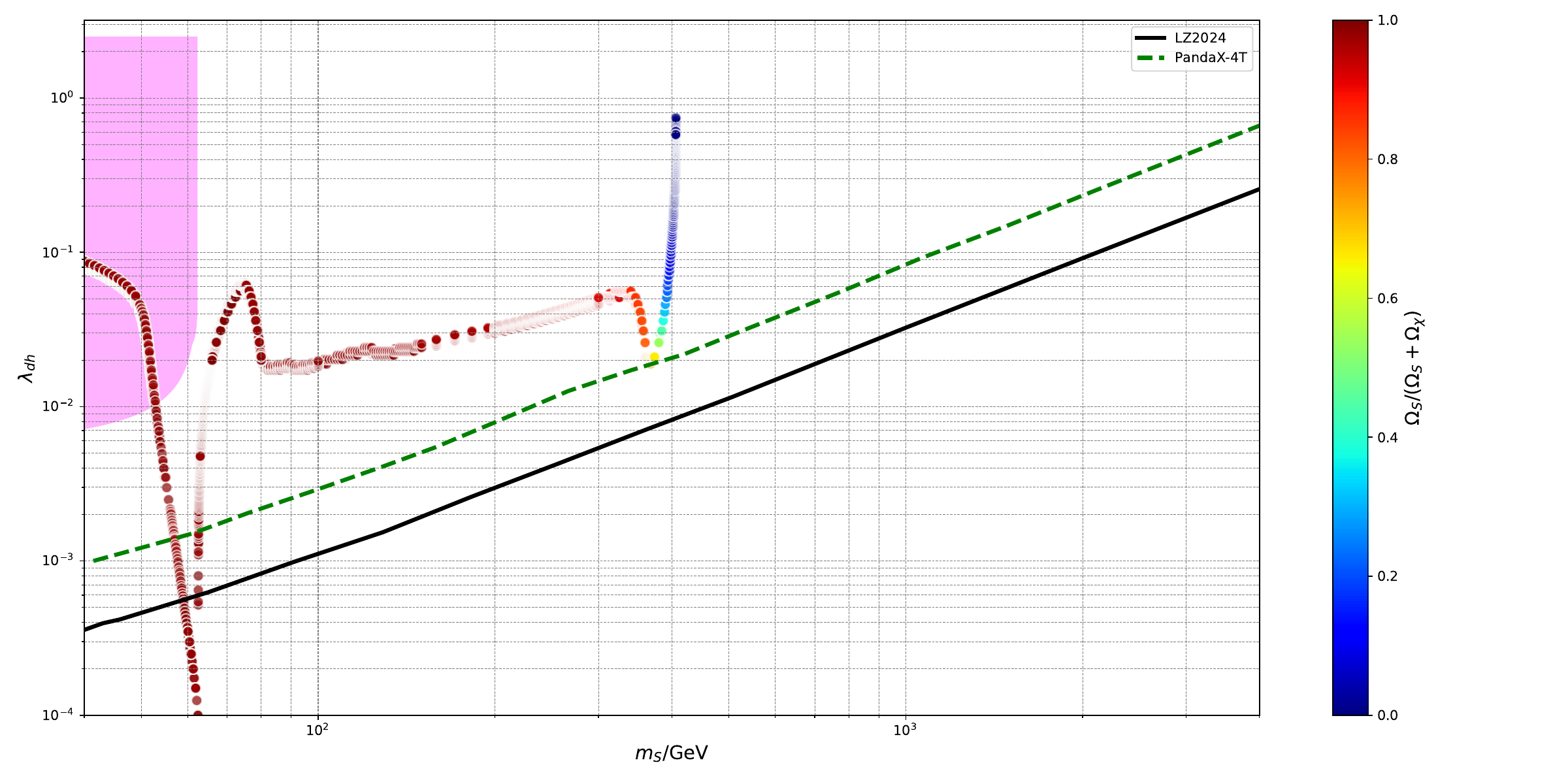}}
\subfigure[]{\includegraphics[height=5.5cm,width=6cm]{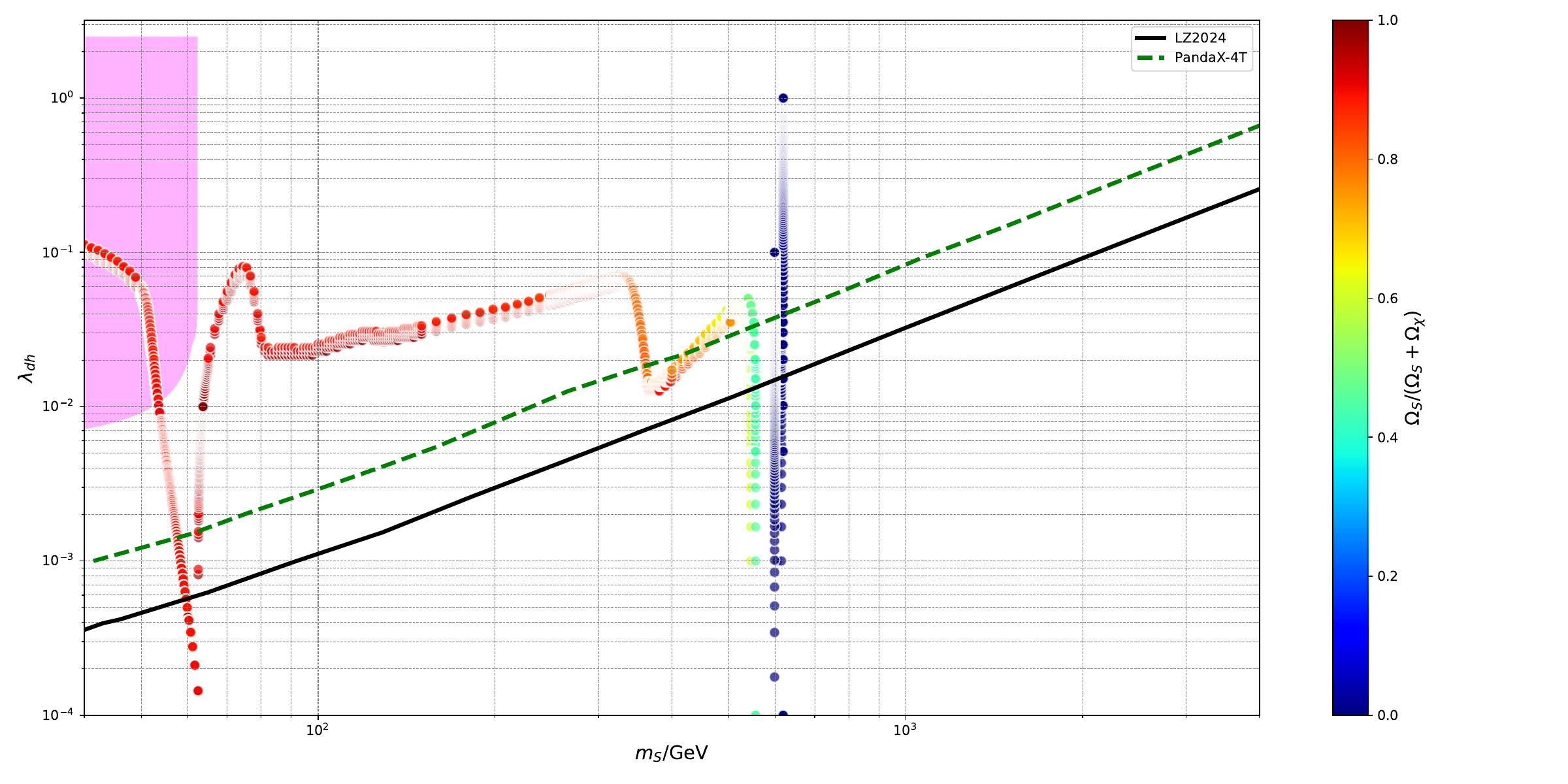}}
\subfigure[]{\includegraphics[height=5.5cm,width=6cm]{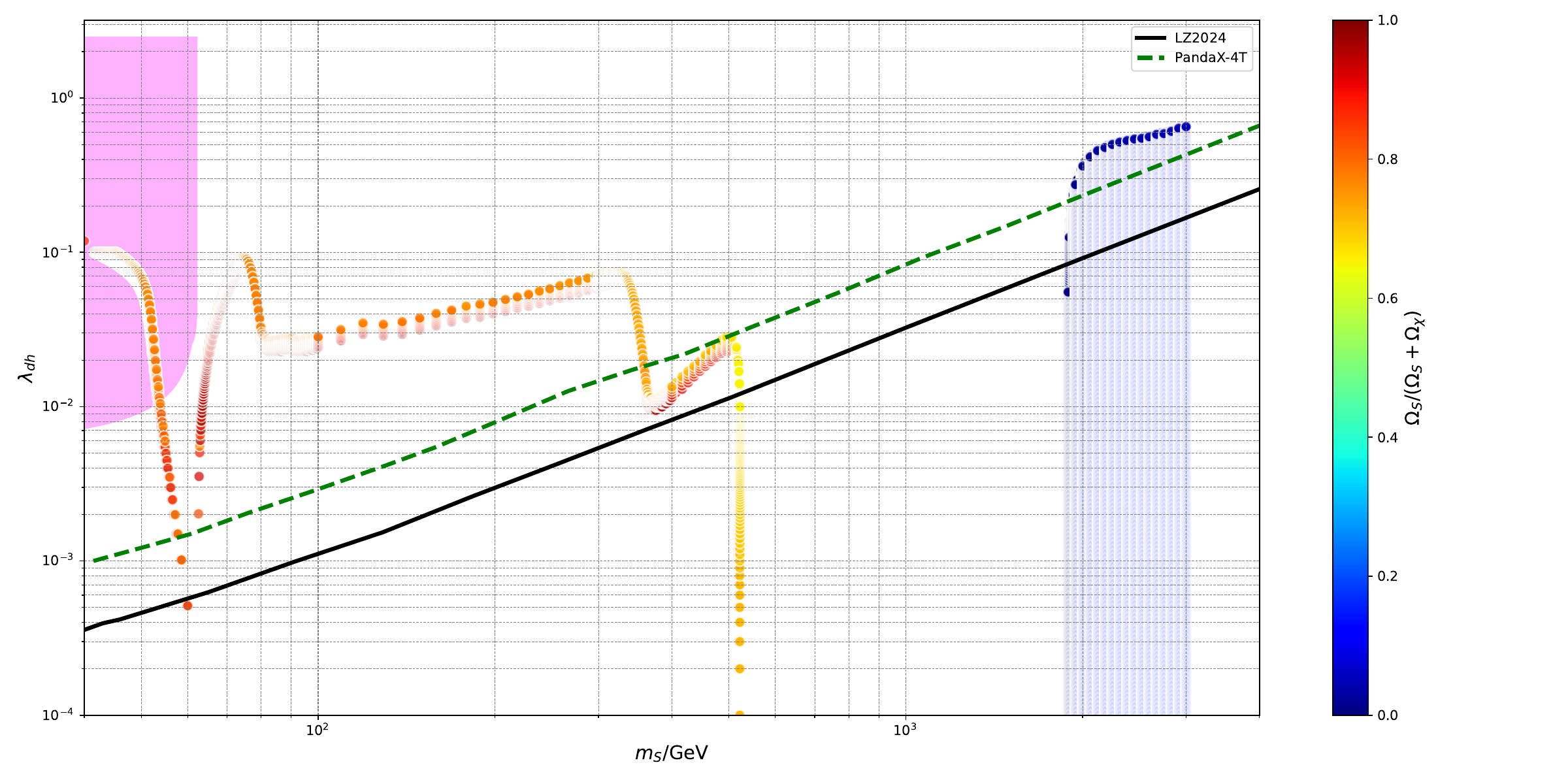}}
\caption{ The combined constraints on the parameter space of $m_S-\lambda_{dh}$, where  we fix $y_{sf}=1$,$\lambda_{ds}=3.14$ and $m_2=600$ GeV. In (a) we set $m_{\chi}=400$ GeV. In (b) we set $m_{\chi}=590$ GeV, and in (c) we set $m_{\chi}=800$ GeV.
Points with diffrent colors satisfying relic density constraint correspond to the fraction $\Omega_S/(\Omega_S+\Omega_{\chi}$, the magenta region is excluded by Higgs invisible decay, the green and black lines respresent the upper bound of $\lambda_{dh}$ for different $m_S$ arising from PandaX-4T \cite{PandaX-4T:2021bab}and latest LZ result \cite{LZ:2024zvo} in the absence of $\chi$.  }
\label{fig8}
\end{figure}
In this part, we show the parameter space under Higgs invisible decay, relic density and direct detection constraint, and the results are given as follows, where we fix $y_{sf}=1$, $\lambda_{ds}=3.14$, $m_2=600$ GeV and vary $m_S$ as well as $\lambda_{sh}$ with $m_S \subseteq [40~ \mathrm{GeV}, 3000~ \mathrm{GeV}]$ and $ \lambda_{sh} \subseteq [10^{-5},3.14]$.
We show the combined constraints on the parameter space of $m_S-\lambda_{dh}$ in Fig.~\ref{fig8}. In Fig.~\ref{fig8}(a) we set $m_{\chi}=400$ GeV. In Fig.~\ref{fig8}(b) we set $m_{\chi}=590$ GeV, and in Fig.~\ref{fig8}(c) we set $m_{\chi}=800$ GeV.
Points with different colors satisfying relic density constraint correspond to the fraction $\Omega_S/(\Omega_S+\Omega_{\chi})$, the magenta region is excluded by Higgs invisible decay, the green and black lines represent the upper bound of $\lambda_{dh}$ for different $m_S$ arising from PandaX-4T \cite{PandaX-4T:2021bab}and latest LZ result \cite{LZ:2024zvo} in the absence of $\chi$. Direct detection results put the most stringent constraint on the parameter space according to Fig.~\ref{fig8}. In Fig.~\ref{fig8}(a) with $m_{\chi}<m_2$, the allowed value for $m_S$ is about $m_1/2$  while another region is excluded by direct detection constraint. In Fig.~\ref{fig8}(b) with $m_{\chi}\approx m_2$, we have three viable parameter space for $m_S$ with  $m_S \approx m_1/2$ , $m_S \approx m_{\chi}$ and $m_S \approx m_2$. For the latter two regions, the t-channel processes of $SS \to \chi\chi$ and $SS \to h_2h_2$ are efficient so that the allowed $\lambda_{dh}$ can be much smaller and not excluded by direct detection constraint. In Fig.~\ref{fig8}(c) with $m_{\chi} > m_2$, we have three viable region for $m_S$ with $m_S \approx m_1/2$, $m_S \approx m_2$ and $m_S > 1890$ GeV. When $m_S$ is larger than 1890 GeV, the processes $SS \to \chi \chi$ as well as $SS \to h_2h_2$  play a dominant role in determining dark matter relic density and one can have a flexible parameter space for  $(m_S,\lambda_{dh})$ under relic density and direct detection constraint. Note that for the three different cases in Fig.~\ref{fig8}, when $m_S$ is small, the fraction $\Omega_S/(\Omega_S+\Omega_{\chi})$ is large and with the increase of $m_S$, the fraction decreases due to the large interaction of $SS \to h_2h_2$ and $SS \to \chi\chi$.
 
\section{summary}

The WIMP DM models are facing serious challenges for the null result of the current direct detection experiments, which put the most stringent constraint on the parameter space of the models.  One solution to alleviate the conflict is the multi-component dark matte model, where the quantity to be compared
against the direct detection limits provided by the experimental collaborations is not the cross-section itself but rather the product of dark matter fraction times the respective cross-section. In this work, we consider a two-component dark matter model under $Z_2 \times Z_4$ symmetry, where a singlet scalar $S$ and a fermion  $\chi$ are introduced as dark matter candidates. In addition, we introduce another new singlet scalar with non-zero vev so that $\chi$ can obtain mass after spontaneously symmetry breaking.  Under the decoupling limit, the fermion dark matter production is just determined by the dark sectors and direct detection constraint will only limit the parameter space of the scalar dark matter.  Provided that the different mass hierarchies of the dark sectors will make a difference in the reaction rate of dark matter annihilation processes, we will have different viable parameter spaces. In this work, we have six different cases with possible mass ordering. We randomly scan a chosen parameter space with the six different cases under relic density constraint. For $\chi$ is the lightest, dark matter production of $\chi$ is determined by the forbidden channels, and we come to the so-called "Forbidden dark matter" region for $\chi$. The allowed value of $m_{\chi}$ as well as $m_S$ can constrained within a narrow mass region with $m_{\chi}$, $m_S$ and $m_2$ are degenerate. For another four cases, the annihilation of $\chi$ to $S$ or $h_2$ is kinetically allowed at zero temperature, and we have a flexible parameter space for these cases.  Moreover, we consider the combined limits arising from Higgs invisible decay, dark matter relic density and direct detection constraints on the parameter space. Within the chosen parameter space, we have three possible allowed mass regions for $m_S$ with $m_s \approx 1/2 m_1$, $m_S \approx m_2$ and $m_S >1890$ GeV,  which depends on the mass hierarchy between $m_{\chi}$ and $m_2$. For $m_{\chi} \geqslant m_2$, the viable parameter space for $(m_S,\lambda_{dh})$  under the combined constraints is wider. 
\begin{acknowledgments}
\noindent
Hao Sun is supported by the National Natural Science Foundation of China (Grant No. 12075043, No. 12147205). XinXin Qi is supported by the National Natural Science Foundation of China Grant No. 12075043.
\end{acknowledgments}
\appendix
\section{Appendix}
\label{appA}
\subsection{Cross-section}
 In this part, we show the cross-section of $\chi \chi \to SS$, $\chi\chi \to h_2 h_2$ and $SS \to h_2h_2$ in the case of decoupling limit. The expressions are given as follows :
 \begin{align}
 \sigma_{\chi\chi \to SS} &=\frac{\lambda_{ds}^2 v_0^2 y_{sf}^2 \sqrt{\frac{\left(s-4 m_S^2\right) \left(s-4 m_{\chi}^2\right)}{s^2}}}{8 \pi \left(m_2^2-s\right)^2}
 \end{align}
\begin{align}
  \sigma_{\chi\chi \to h_2h_2} &=\frac{y_{sf}^2}{8\pi s (s-4 m_{\chi}^2)} ( (\frac{24 \sqrt{2} m_2^2 m_{\chi} y_{sf}}{m_2^2 v_0-s v_0}+\frac{9 m_2^4 (s-4 m_{\chi}^2)}{v_0^2 (m_2^2-s)^2} 
  -\frac{4 y_{sf}^2 (3m_2^4-16m_2^2 m_{\chi}^2+2 m_{\chi}^2 (8 m_{\chi}^2+s))}{m_2^4-4 m_2^2 m_{\chi}^2+m_{\chi}^2 s})\notag\\
  &\frac{\sqrt{(s-4 m_2^2) (s-4 m_{\chi}^2)} }{4}
  +\frac{y_{sf}} {2 m_2^4-3 m_2^2 s+s^2}\log (\frac{-\sqrt{(s-4 m_2^2) (s-4m_{\chi}^2)}-2m_2^2+s}{\sqrt{(s-4m_2^2) (s-4 m_{\chi}^2)}-2 m_2^2+s}) (y_{sf} (m_2^2-s) \notag\\
 & (6 m_2^4-4 m_2^2 (4 m_{\chi}^2+s)-32 m_{\chi}^4+16m_{\chi}^2 s+s^2)-\frac{3 \sqrt{2} m_2^2 m_{\chi}(2 m_2^2-s) (2 m_2^2-8 m_{\chi}^2+s)}{v_0}))
 \end{align}
 \begin{align}
 \sigma_{SS \to h_2h_2} &=\frac{\lambda_{ds}^2}{8 \pi s (s-4 m_S^2)} (\frac{8 \lambda_{ds}^2 v_0^4 \log (\frac{\sqrt{(s-4 m_2^2) (s-4 m_{\chi}^2)}+6 m_{\chi}^2-4 m_S^2-s}{-\sqrt{(s-4 m_2^2) (s-4 m_{\chi}^2)}+6m_{\chi}^2-4 m_S^2-s})}{m_2^2+3 m_{\chi}^2-2 m_S^2-s}+\sqrt{(s-4 m_2^2) (s-4 m_{\chi}^2)}\notag\\
 &(\frac{8 \lambda_{ds}^2 v_0^4}{m_2^4-4 m_2^2 m_{\chi}^2+m_{\chi}^2 s}+\frac{(2 m_2^2+s)^2}{(m_2^2-s)^2})+8\lambda_{ds} v_0^2 \log (\frac{\sqrt{(s-4 m_2^2)(s-4 m_{\chi}^2)}+2 m_2^2-s}{-\sqrt{(s-4 m_2^2) (s-4 m_{\chi}^2)}+2 m_2^2-s}) \notag\\
& (\frac{\lambda_{ds} v_0^2}{m_2^2+3 m_{\chi}^2-2 m_S^2-s}+\frac{3 m_2^2}{s-m_2^2}+1))
 \end{align}
where $s$ is the squared center-of-mass energy, $\chi\chi \to SS$ is a $2 \to 2$ scattering process  mediated by $h_2$. The process $\chi\chi \to h_2h_2$ involves $\chi$-mediated channel, as well as  $h_2$-mediated channel, and the process of $SS \to h_2h_2$, involves $h_2$-mediated channel as well as the $2\to2$ scattering.
\subsection{Forbidden dark matter}
When $\chi$ is the lightest among the dark sectors, $\chi$ production is determined by forbidden channels, which are kinetically forbidden at zero temperature.  At high temperatures, the thermally averaged forbidden annihilation rates are:
\begin{align}
<\sigma_{\chi S} v>&=<\sigma_{S\chi}v>\frac{n^2_{Seq}}{n^2_{\chi eq}}\approx <\sigma S\chi>e^{-2\Delta_S x},\\
<\sigma_{\chi h_2} v>&=<\sigma_{h_2\chi}v>\frac{n^2_{h_2eq}}{n^2_{\chi eq}}\approx <\sigma S\chi>e^{-2\Delta_h x}.
\end{align}
where $x\equiv m_{\chi}/T$, $\Delta_S \equiv (m_S-m_{\chi})/m_{\chi}$, $\Delta_h\equiv(m_2-m_{\chi})/m_{\chi}$, $\sigma_{\chi S} \equiv \sigma_{\chi\chi \to SS}$, $\sigma_{\chi h_2} \equiv \sigma_{\chi\chi \to h_2h_2}$, and  $\sigma_{S\chi}$ as well as $\sigma_{h_2\chi}$ are  the cross-section for the inverse processes. $n_{\chi eq},n_{Seq}$ and $n_{h_2 eq}$ represent  number density of $\chi, S$ and $h_2$ at thermally equilibrium.
\begin{figure}[htbp]
\centering
 \subfigure[]{\includegraphics[height=5.5cm,width=6cm]{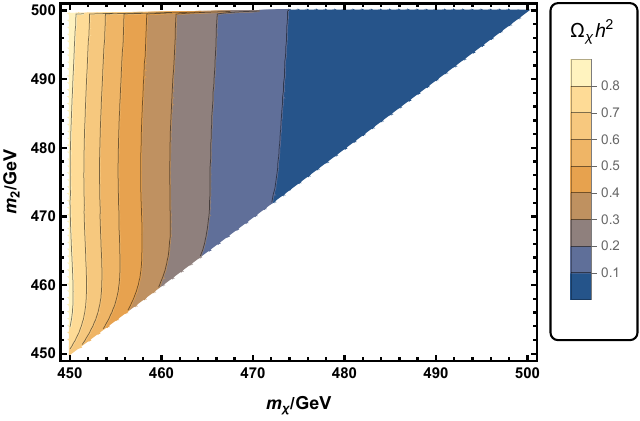}}
\subfigure[]{\includegraphics[height=5.5cm,width=6cm]{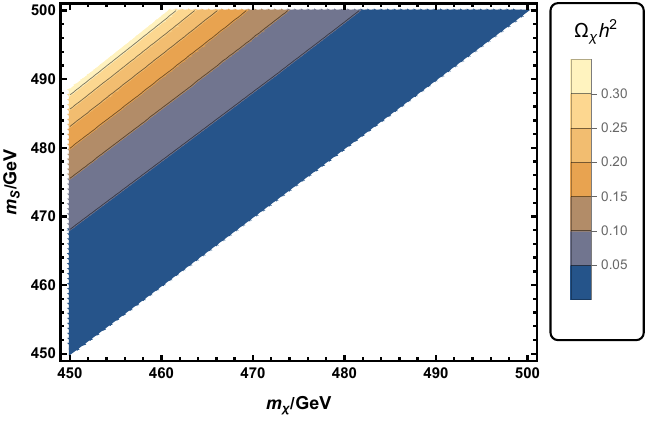}}
\caption{Contour plots of $m_{\chi}-m_2$ (a) in the case of  $m_{\chi} <m_S<m_2$ (a) and $m_{\chi} <m_2<m_S$(b), where the legends represent $\Omega_{\chi}h^2$. }
\label{fig9}
\end{figure}

We have two cases with $m_{\chi} <m_S<m_2$ and $m_{\chi}<m_2<m_S$, and we give the results in Fig.~\ref{fig9}(a) and Fig.~\ref{fig9}(b) respectively, where we fixed $y_{sf}=3.14,\lambda_{ds}=0.03$ and $\lambda_{dh}=1.5$.  In  Fig.~\ref{fig9}(a), we show the
Contour plots of $m_{\chi}-m_2$, where we set $m_S=500$ GeV. For a fixed $m_{\chi}$, the value of $m_2$ makes little difference in $\Omega_{\chi}h^2$, and with the increase of $m_{\chi}$, the processes of  $\chi \chi \to SS$ as well as $\chi\chi \to h_2 h_2$ become more efficient so that $\Omega_{\chi}h^2$ decrease. Particularly,  as $m_{\chi} \approx m_S$ and $\Delta_S \approx 0$ the value of $\Omega_{\chi}h^2$ is almost unchanged regardless of $m_{\chi}$. In Fig.~\ref{fig9} (b), we set  $m_2=500$ GeV and vary $m_{\chi}$ as well as $m_S$. For a fixed $m_{\chi}$, $\Omega_{\chi}h^2$ increases with the increase of $m_S$, and the upper bound of $m_S$ is constrained to guarantee the efficiency of the process of $\chi \chi \to SS$. With the increase of $m_{\chi}$, the value of $\Omega_{\chi}h^2$  decrease as the case of $m_{\chi} <m_S<m_2$, and the value of $m_S$ is more flexible when $m_{\chi}$ is larger than about $462$ GeV.


\bibliographystyle{JHEP}
\bibliography{v1.bib}

\end{document}